\documentclass[10pt,compsoc,journal,twoside]{IEEEtran}

\usepackage{amsmath}
\usepackage{graphicx}
\usepackage{epstopdf}
\usepackage{array}
\usepackage{url}
\usepackage{multirow}
\usepackage{dblfloatfix}
\usepackage{cite}\usepackage{bm}
\usepackage{eurosym}
\usepackage{balance}
\DeclareMathOperator*{\argmax}{argmax}

\usepackage{ragged2e}
\usepackage{xcolor}

\newif\ifproofread

\newcommand{\changemarker}[1]{%
	\ifproofread
	\textcolor{red}{#1}%
	\else
	#1%
	\fi
}

\proofreadfalse

\begin{document}
%
\title{AMIGOS: A Dataset for Affect, Personality and Mood Research on Individuals and Groups}
%
%
%

\author{Juan~Abdon~Miranda-Correa,~\IEEEmembership{Student Member,~IEEE,}
        Mojtaba~Khomami~Abadi,~\IEEEmembership{Student Member,~IEEE,}
        Nicu~Sebe,~\IEEEmembership{Senior Member,~IEEE,}
        and~Ioannis~Patras,~\IEEEmembership{Senior Member,~IEEE}
\thanks{Juan~Abdon~Miranda-Correa and Ioannis~Patras are with the School of Computer Science and Electronic Engineering, Queen Mary University of London, UK. E-mail: \{j.a.mirandacorrea,i.patras\}@qmul.ac.uk.}
\thanks{Mojtaba~Khomami~Abadi and Nicu~Sebe are with the Department of Information Engineering and Computer Science, University of Trento, Italy. E-mail: \{khomamiabadi,sebe\}@disi.unitn.it.}
}

%
%

\markboth{IEEE TRANSACTIONS ON AFFECTIVE COMPUTING, VOL.~?, NO.~?, ?~?}%
{MIRANDA-CORREA ET AL: AMIGOS: A dataset for Mood, personality and affect research on Individuals and GrOupS}
%





\IEEEtitleabstractindextext{%
\begin{justify}
\begin{abstract} 
	\changemarker{We present AMIGOS-- A dataset for Multimodal research of affect, personality traits and mood on Individuals and GrOupS. Different to other databases, we elicited affect using both short and long videos in two social contexts, one with individual viewers and one with groups of viewers. The database allows the multimodal study of the affective responses, by means of neuro-physiological signals of individuals in relation to their personality and mood, and with respect to the social context and videos' duration. The data is collected in two experimental settings. In the first one, 40 participants watched 16 short emotional videos. In the second one, the participants watched 4 long videos, some of them alone and the rest in groups. The participants' signals, namely, Electroencephalogram (EEG), Electrocardiogram (ECG) and Galvanic Skin Response (GSR), were recorded using wearable sensors. Participants' frontal HD video and both RGB and depth full body videos were also recorded. Participants emotions have been annotated with both self-assessment of affective levels (valence, arousal, control, familiarity, liking and basic emotions) felt during the videos as well as external-assessment of levels of valence and arousal. We present a detailed correlation analysis of the different dimensions as well as baseline methods and results for single-trial classification of valence and arousal, personality traits, mood and social context. The database is made publicly available.}

\end{abstract}
\end{justify}

}

\maketitle
\IEEEpeerreviewmaketitle

%
\IEEEpeerreviewmaketitle

\section{Introduction}\label{Introduction}
	Affective computing aims for the detection, modeling and synthesis of human emotional cues in Human-Computer Interaction \cite{koelstra2012deap}. In this field, an increasing interest has arisen for considering the user's affective responses when making computational decisions. For instance, Chanel et al \cite{ChanelRBP11} modified the difficulty of a video game according to the user's affective (emotional) state to maintain high engagement. 
	In a hypothetical scenario, the time-line of a movie could be adapted to elicit specific affective states, taking into account factors such as the viewer's predicted emotions, personality and mood. Hence, in these scenarios, it is very important to reliably predict such factors. 
	
	Advances on the prediction of affective states have been boosted by the availability of annotated affective databases, which act as benchmark for many researchers to develop their methodologies. These databases have used stimuli, such as music videos \cite{koelstra2012deap}, short videos \cite{DECAF2015,koelstra2009eeg}, and diverse emotion elicitation methods \cite{zhang2016multimodal}. They include information from different modalities (e.g. EEG, facial expression).  
	
	Available multimodal affective databases have focused on the study of affective responses of participants in individual \cite{koelstra2012deap,SoleymaniMAHNOB2012}, or pairs of people/limited agent settings \cite{McKeown2012}. However, in real life, affective experiences are often performed in social contexts (e.g. movies and games are commonly engaged by groups of people together). In such contexts, the individual experiences do not depend only on the user and the content, but also on the implicit and explicit interactions that can occur between the personalities, reactions, moods and emotions of other group members. Additionally, different aspects of affect and personality could be inhibited or amplified depending on the social context of a person. Therefore, current databases have ignored an important dimension for the study of affect.
	
	Databases for personality research have considered information related to linguistics in written text \cite{Wilson1988}, social networks activity \cite{kosinskiFacebook2015}, and behavior in group activities \cite{Pianesi2007}. However they have largely ignored the study of both, affect and personality, through the use of physiological signals, which have shown to carry valuable information for personality recognition \cite{FG_Paper,Wache2015}. 
	
	Therefore, there is a need of multimodal databases for the study of people's emotions, personality and mood, with subjects in both alone and group settings. The multimodal framework would benefit from the inclusion of neurological and peripheral physiological signals.
	
	Our contribution to the field is A dataset for Multimodal research of affect, personality traits and mood on Individuals and GrOupS (AMIGOS) by means of neuro-physiological signals. The dataset consists of multimodal recordings of participants and their responses to emotional fragments of movies. In our dataset: 
	(i) The participants took part in two experiments. In each of them, the participants watched one of two sets of stimuli, one of short videos and one of long videos, while their implicit responses, namely, Electroencephalogram (EEG), Electrocardiogram (ECG), Galvanic Skin Response (GSR), frontal HD video, and both RGB and depth full body videos were recorded. The recordings have been precisely synchronized to allow the study of affective responses, personality and mood from the different modalities simultaneously.
	(ii) In the first experiment, all participants watched the set of short videos in individual setting. In the second experiment, some of the participants took part in individual setting and some of them in group settings. Then they watched the set of long videos. 
	(iii) The participants have been profiled according to their personality through the Big-Five personality traits model, and according to their mood through the Positive Affect and Negative Affect Schedules (PANAS). 
	(iv) Affective annotation has been obtained with both internal and external annotations. In the internal annotation, participants performed self-assessment of their affective levels at the beginning of each experiment and immediately after each video. In the external annotation, the recordings of both sets of videos were off-line annotated by 3 annotators on both valence and arousal scales, using a method that allows the direct comparison of the affective responses from both experiments.  
	(v) The physiological signals have been recorded using commercial wearable sensors that allow more freedom for the participants than conventional laboratory equipment (e.g. Biosemi ActiveTwo\footnote{http://www.biosemi.com/}) used in \cite{koelstra2012deap,SoleymaniMAHNOB2012,DECAF2015} and of better quality than the equipment used in \cite{Wache2015}. 
	The database is available to the academic community\footnote{http://www.eecs.qmul.ac.uk/mmv/datasets/amigos/}.
	
	In this work, we present a comparison between the internal and external annotations of valence and arousal. We then perform a detailed correlation analysis between the affective responses elicited by the short and long videos with respect of social context (whether a participant was alone or in a group during the experiment) and between the participants' personality traits, PANAS and social context. We also present baseline methodologies and results for single-trial prediction of valence and arousal, and for prediction of personality traits, PANAS and social context, using neuro-physiological signals (EEG, ECG and GSR) as single modalities and fusion of them.
	
	Our main findings are as follows: 
	(i) We show that there is significant correlation between the internal and external annotation of valence and arousal for the short videos experiment, which indicates that external annotation is a good predictor of the affective state of participants. 
	(ii) We show, by correlation analysis of the external annotations, that in the eyes of the annotators, participants seem to have low arousal in low valence moments and high arousal for high valence moments. 
	(iii) \changemarker{We found significant differences in the distribution of valence and arousal, externally annotated, between the participants that were alone compared to the participants that were in groups during the long videos experiment. It was different for the short videos experiment where the distribution of arousal and valence for the 2 sets of participants are not statistically different ($p>0.05$). This result was expected since, as stated before, all the participants watched the short videos within the same social context (alone).} 
	(iv) We found significant negative correlations between the scores of negative affect (NA) and the ones of extraversion, agreeableness, emotional stability and openness, and significant positive correlations between the scores of agreeableness and both extraversion and positive affect (PA), between consciousness and emotional stability, and between PA and arousal. 
	\changemarker{Finally, (v) our method for personality traits, mood and social context prediction based on neuro-physiological signals of short and long videos outperforms a previous study \cite{FG_Paper} in prediction of extroversion, emotional stability, PA and NA usign EEG and in prediction of conscientiousness, openness and conscientiousness using physiological signals (ECG and GSR)}.
	
	In section~\ref{RelatedWorks}, works related to the modeling and assessment of affect, personality and mood are discussed, and a survey of the main multimodal databases available for affect and personality research and a comparison with our are presented.
	Section~\ref{Modalities_Equipment} presents the experimental scenarios, stimuli selection, modalities and equipment used to record the implicit responses. Then, an overview of the experimental setup for both experiments and the methods employed for assessment of affect, personality traits and mood (PANAS) are described. 
	In Section~\ref{Analysis}, the data obtained from the different experiments is analyzed. 
	Section~\ref{Evaluation} presents our method for single trial valence and arousal recognition as well as our approach for personality traits, PANAS and social context recognition using neuro-physiological signals. The results are then presented and discussed. 
	Finally, we conclude in section~\ref{conclusions}.	
\section{Related Works}\label{RelatedWorks}
	In this section, we make a review of the works related with modeling and assessment of affect, personality and mood. Next, we make a review of important databases that study affect, personality and mood.
	\subsection{Affect, Personality and Mood}
	Plutchnik \cite{Plutchik2001} has defined emotion as a complex chain of loosely connected events that begins with a stimulus and includes feelings, psychological changes, impulses to action and specific, goal-directed behavior. The most common approaches to model affect are categorical and dimensional. The first approach claims that there exists a small number of emotions that are basic and recognized universally; The most common of these models is the Six Basic Emotions model, presented by Ekman et al \cite{Ekman1975utf}, that categorizes emotions into fear, anger, disgust, sadness, happiness and surprise. The dimensional approach considers that affective states are inter-related in a systematic way (\textit{e.g.} the Plutchik's emotion wheel \cite{Plutchik2001}). Russell \cite{Russell1980} introduced the Circumplex Model of Affect, where affective states are represented in a two dimensional space with arousal (the degree an emotion feels active) and valence (the degree an emotion feels pleasant) as the main dimensions.
	
	Affective experiences are also modulated by people's internal factors, such as mood and personality \cite{Chevalier2015}. Personality refers to stable individual characteristics, that explain and predict behavior \cite{matthews2003personality}. The Big-Five factor model \cite{Perugini2002} describes personality in terms of five traits (dimensions) namely Extraversion (sociable vs reserved), Agreeableness (compassionate vs dispassionate and suspicious), Conscientiousness (dutiful vs easy-going), Emotional stability (nervous vs confident) and Openness to experience (curious vs cautious). 
	The common method to measure these dimensions is the use of questionnaires such as the Neuroticism, Extraversion and Openess Five Factor Inventory (NEO-FFI) \cite{Costa1992} and the Big-Five Marker scale (BFMS) \cite{Perugini2002}.
	
	Mood refers to baseline levels of affect that define people´s experiences. It is commonly modeled using the two dimensions called Positive Affect (PA) and Negative Affect (NA) scales \cite{Watson1985}. PA and NA are related to corresponding affective trait dimensions of positive and negative emotionality \cite{Watson1985}. PA reflects the extent to which a person feels enthusiastic, active and alert. 
	In contrast, NA is a general dimension of subjective distress and unpleasant engagement.
	In order to measure these two dimensions (PA and NA), Watson et al \cite{Watson1988} developed the Positive and Negative Affect Schedules (PANAS) that consist of two 10-item mood scales; These schedules have shown to be internally consistent, uncorrelated and stable over a 2-month time period.

	\subsection{Databases for Affective Computing}
	Databases for the study of affective computing have been developed to allow researchers to compare results. Here, we will review databases based on video, neurological signals and/or physiological signals modalities. As far as we know, there is not a single database developed for mood research. 
	
	Databases for the study of affect recognition based on video have focused mainly on the analysis of facial expressions. One of the main examples is the Sustained Emotionally Colored Machine-human Interaction using Nonverbal Expression (SEMAINE) database \cite{McKeown2012}. It consists of high-quality, multimodal recordings of 150 participants in emotionally colored conversations. It is annotated for valence, arousal and Facial Action Coding System (FACS) action units (AUs). Another example is the Affectiva-MIT Facial Expression Dataset (AM-FED) \cite{McDuffKSACP13}. It is a labeled dataset of spontaneous facial responses recorded in natural settings on the Internet. The dataset consists of 242 facial videos, labels of the presence of 10 symmetrical and 4 asymmetrical AUs, 2 head movements, smile, general expressiveness, feature tracker fails, gender, location of 22 automatically detected landmark points and self-report responses of familiarity, liking and desire to watch again. The Denver Intensity of Spontaneous Facial Action (DISFA) database \cite{Mavadati2013} consists of labeled stereo video recordings of 27 adults while watching a video clip. Labels consist of presence, absence and intensity of 12 facial AUs.
	
	Databases for affect research based on physiological signals include the MAHNOB-HCI \cite{SoleymaniMAHNOB2012}. It is a multimodal database that consists of synchronized recordings of face video, audio signals, eye gaze data and physiological signals (ECG, GSR, respiration amplitude (RA), skin temperature (ST) and EEG) of 27 participants while watching first, 20 videos, and second, short videos and images with relevant/non-relevant tags. It includes the self-reports of the felt emotions using arousal, valence, dominance, predictability scales, emotional keywords and agreement or disagreement with the tags. Koelstra et al present the DEAP database \cite{koelstra2012deap}, with the purpose of implicit affective tagging from EEG and peripheral physiological signals (GSR, RA, ST, ECG, blood volume, Zygomaticus and Trapezius muscles Electromyogram and Electrooculogram) research. It consists of video and signals' recordings of 32 participants while watching 40 music video clips. It includes self-assessment of arousal, valence, liking, dominance and familiarity. A similar database that uses Magnetoencephalogram (MEG) is the DECAF database, which includes recordings of 30 participants in response to 40 one-minute music video and 36 movie clips. More recently, Zhang et al \cite{zhang2016multimodal} collected the Multimodal Spontaneous Emotion Corpus for Human Behavior Analysis. It includes 140 participants from various ethnic origins. They used 10 different emotion elicitation methods for specific target emotions (ee.gg. surprise, disgust, fear). Recorded signals are 3D and 2D videos, thermal sensing, electrical conductivity of the skin, respiration, blood pressure and hearth rate. It includes annotations of the occurrence and intensity of AUs. These databases have not considered studying participants in group setting.
	
	One of the first databases for personality research using video modality, is the Mission Survival II corpus \cite{Pianesi2007}. It is a multimodal annotated collection of video and audio recordings (using 4 cameras and 17 microphones) of four meetings, of 4 participants engaging in a mission survival task. Participants were profiled in terms of the Ten Item Personality Inventory \cite{Gosling2003} to account for their personality states (moments where participants act more or less introvert/extravert, creative, ect). This dataset is not intended for affect research. A recent multi-modal database for implicit personality and affect recognition is the ASCERTAIN \cite{Subramanian2016}. It includes recordings of the EEG, ECG, GSR and facial video of 58 users, while viewing short movie clips. They showed that 
	personality differences are better revealed while comparing user responses to emotionally homogeneous videos (videos that share the same quadrant of the valence-arousal space). This database only includes participants in individual configuration and does not share data about mood of participants. 
	
	To the best of our knowledge there are not databases for personality research based on neurological or physiological signals and that studies participants in both individual and group settings. In Table~\ref{Databases}, we summarize the characteristics of the reviewed databases and compare them to ours. 

	\begin{table*}[t]
		\centering
		\fontsize{8}{8}\selectfont
		\renewcommand{\arraystretch}{1.2}
		\caption{\label{Databases} Summary of characteristics of databases for affect and personality. Last row is our database.}
		\scalebox{0.82}[0.88]{
			\begin{tabular}{|p{.135\paperwidth}|p{0.030\paperwidth}|p{0.070\paperwidth}|p{0.175\paperwidth}|p{0.220\paperwidth}|p{0.270\paperwidth}|}
			\hline
			\hline
			\centering{\textbf{Database}}	& \centering{\textbf{No. Part.}} & \centering{\textbf{Individual vs. Group}} & \centering{\textbf{Purpose}} & \centering{\textbf{Modalities}} & \textbf{Annotations} \\ 
			\hline
			\centering{\textbf{SEMAINE }\cite{McKeown2012}}	& 150 & Individual & Emotion recognition based on facial expressions & Audio and Visual  & Valence, arousal and FACS. \\ 
			\hline
			\centering{\textbf{AM-FED } \cite{McDuffKSACP13}} & 242  & Individual & Spontaneous facial expression recognition "In-the-Wild" & Visual & 14 AUs, 2 head movements, smile, expressiveness and 22 landmark points. Self-assessment of familiarity, liking and desire to watch again. \\ 
			\hline
			\centering{\textbf{DISFA } \cite{Mavadati2013}}	& 27 & Individual & Spontaneous facial action recognition & Visual & 12 AUs. \\ 
			\hline
			\centering{\textbf{MAHNOB-HCI} \cite{SoleymaniMAHNOB2012}}	& 27 & Individual & Emotion recognition and implicit tagging & Visual, Audio, Eye Gaze, ECG, GSR, Respiration Amplitude, Skin temperature, EEG & Self-assessment of valence, dominance, predictability and emotional keywords. Agreement/disagreement with tags. \\ 
			\hline
			\centering{\textbf{DEAP} \cite{koelstra2012deap}} & 32 & Individual & Implicit affective tagging from EEG and peripheral physiological signals & EEG, GSR, Respiration Amplitude, Skin Temperature, Blood Volume, Electromyogram and Electrooculogram. Visual for 22 participants. & Self-assessment of arousal, valence, liking, dominance and familiarity.\\ 
			\hline
			\centering{\textbf{DECAF} \cite{DECAF2015}} & 30 & Individual & Affect recognition & MEG, Near-infra-red facial video, horizontal Electrooculogram, ECG and trapezius-Electromyogram. & Self-assessment of valence, arousal and dominance. Continuous annotation of valence and arousal of the stimuli.\\ 
			\hline
			\centering{\textbf{Zhang et al corpus} \cite{zhang2016multimodal}} & 140 & Individual & Emotional behaviour research & 3D dynamic imaging, Visual, Thermal sensing, EDA, Respiration, Blood Pressure and Hearth Rate & Occurrence and intensity of AUs. Features from 3D, 2D and Infra-red sensors.   \\ 
			\hline
			\centering{\textbf{Mission Survival II}\cite{Pianesi2007}} & 16 & 4 people group & Personality states research & Audio and Visual & Personality states by the Ten Item Personality Inventory.  \\ 
			\hline
			\centering{\textbf{ASCERTAIN}  \cite{Subramanian2016}} & 58 & Individual & Personality and Affect & EEG, ECG, GSR and Visual & Big-Five personality traits, self-assessment of valence and arousal.   \\ 
			\hline
			\centering{\textbf{AMIGOS}}	& 40 & Individual \& 4 people group & Affect, personality, mood and social context recognition  & Audio, Visual, Depth, EEG, GSR and ECG & Big-Five personality traits and PANAS. Self-assessment of valence, arousal, dominance, liking, familiarity and basic emotions. External annotation of valence and arousal.  \\ 
			\hline
		\end{tabular}
	}
\vspace{-.3cm}
	\end{table*}

\section{Experimental Setup}\label{Modalities_Equipment}
	In this section, the experimental scenarios are described. Then the process followed for the selection of stimuli is explained, and the modalities and equipment used are presented. Then, the experimental protocol is described in detail. Finally, the procedures for internal and external annotation of affect and for participants' personality and mood assessment are introduced.

	\subsection{Experimental scenarios}
		The main objective of this work is to study the personality, mood and affective responses of people engaging with multimedia content in two social contexts, (i) when they are alone (individual setting), and (ii) when they are part of an audience (group setting). At the same time, we study people's affective response to two types of eliciting content. The first type consists of short emotional videos (duration$<$250s) selected to elicit specific affective states in the participants. The second type consists of long videos (duration$>$14min), that present situations that could elicit various affective states over their duration and where the story and the narrative could give context to the affective responses. Therefore, we have designed two experiments, in the first one (Short videos experiment), all participants watched short affective videos in individual setting. In the second experiment (Long videos experiment), the same participants watched long videos, but this time some of them did it in individual setting, while the others did it in group setting.

	\subsection{Stimuli selection}
		Emotion elicitation depends greatly on a careful selection of the stimuli, which needs to be suitable for the objective of the study and allow for consistent results among trials \cite{koelstra2012deap}. In this work, we selected two sets of videos for emotion elicitation. The first one consists of short emotional videos and the second one of long videos. For the first set, 72 volunteers annotated, on the valence and arousal dimensions, the set of 36 videos used in \cite{DECAF2015}. We then classified each of the videos into one of four quadrants of the valence-arousal (VA) space, namely HVHA, HVLA, LVHA and LVLA (H, L, A and V stand for high, low, arousal and valence respectively). From each quadrant, we selected the three videos that lay further to the origin of the scale, totaling 12 videos. Additionally, from the videos used in \cite{SoleymaniMAHNOB2012}, we selected four videos, each corresponding to one of the four quadrants. The total number of selected short videos is 16, 4 for each quadrant of the VA space. We have preserved the IDs used in the original datasets. The selected short videos (51-150s long, $\mu=86.7$, $\sigma=27.8$) with their corresponding category on the VA space and their IDs are listed in Table~\ref{VideoList}.

		\begin{table}[t]
			\centering
			\fontsize{8}{8}\selectfont
			\renewcommand{\arraystretch}{1.2}
			\caption{\label{VideoList} The short videos listed with their sources (Video IDs are stated in parentheses). In the category column, H, L, A and V stand for high, low, arousal and valence respectively.}
			\scalebox{0.82}[0.88]{
				\begin{tabular}{|p{.15\linewidth}|p{0.95\linewidth}|}
				\hline
				\hline
				\small
				\centering{\textbf{Category}}	&  \textbf{Excerpt's source}\\
				\hline
				\centering{\textbf{HAHV}}		& Airplane (4), When Harry Met Sally (5), Hot Shots (9), Love Actually (80)\\  \hline
				\centering{\textbf{LAHV}}		& August Rush (10), Love Actually (13), House of Flying Daggers (18), Mr Beans' Holiday (58)\\  \hline
				\centering{\textbf{LALV}}		& Exorcist (19), My girl (20), My Bodyguard (23), The Thin Red Line (138)\\  \hline
				\centering{\textbf{HALV}}		& Silent Hill (30), Prestige (31), Pink Flamingos (34), Black Swan (36) \\  \hline
			\end{tabular}
		}
	\vspace{-.2cm}
		\end{table}
		
		For the second set of videos, we initially selected 8 video extracts from movies based on their score in the IMDb Top Rated Movies list\footnote{ http://www.imdb.com/chart/top}. We selected movies that could allow us to extract a long segment ($\approx$ 20min) which could be self-contained, did not require previous knowledge from the participants to be understood and with strongly affective multimedia content (good combination of music and colors \cite{Soleymani2008}). Four researchers classified them as belonging to one or more quadrants of the VA space. Finally, 4 videos were selected favoring the extracts that could evoke emotions in different quadrants of the VA space, and making sure all the quadrants were covered. 
		The selected long videos (14.1-23.58min, $\mu=20.0$, $\sigma=4.5$) with their corresponding video ID, source and duration are listed in Table~\ref{VideoList2}.
		
		\def\perc{0.87}
		\begin{table}[t]
			\centering
			\fontsize{8}{8}\selectfont
			\renewcommand{\arraystretch}{1.2}
			\caption{\label{VideoList2} Selected Long Videos with Their ID, Source (Movie title.
				Director. Producer company. Released Year.) and Excerpt Duration.}
			\scalebox{0.82}[0.88]{
					\begin{tabular}{|c|p{\perc\linewidth}|c|}
				\hline
				\hline
				\small
				\textbf{ID}&\centering{\textbf{Source}}&\textbf{Duration}\\
				\hline
				N1 &\multicolumn{1}{m{\perc\linewidth}|}{The Descent. Dir. Neil Marshall. Lionsgate. 2005.}  & 23:35.0\\  \hline
				P1 & \multicolumn{1}{m{\perc\linewidth}|}{Back to School Mr. Bean. Dir. John Birkin. Tiger Aspect Productions. 1994.}  & 18:43.0\\  \hline
				B1 & \multicolumn{1}{m{\perc\linewidth}|}{The Dark Knight. Dir. Christopher Nolan. Warner Bross. 2008.} & 23:30.0\\  \hline
				U1 &\multicolumn{1}{m{\perc\linewidth}|}{Up. Dirs. Pete Docter and Bob Peterson. Walt Disney Pictures and Pixar Animation Studios. 2009.}  & 14:06.0\\  \hline
			\end{tabular}
		  }
	  \vspace{-.2cm}
		\end{table}

	\subsection{Neuro-Physiological Signals and Instruments}\label{signals}
		We recorded three main neural and peripheral physiological signals namely Electroencephalogram (EEG), Electrocardiogram (ECG) and Galvanic Skin Response (GSR), which have shown good performance in affect estimation studies \cite{Bong2012,Ahmad2015wave,Lang1990}. Below we give an introduction of each of them. 
		
		\textbf{EEG}\label{EEG_modality}: Electroencephalogram is a recording of the electrical activity along the scalp. It measures voltage fluctuations resulting from ionic current flows within the brain \cite{friedman2015encyclopedia}.
		EEG signals carry valuable information about the person's affective state \cite{Damasio2000,koelstra2012deap}.
		
		\textbf{GSR}\label{GSR_modality}: Galvanic skin response, also known as electrodermal activity (EDA), measures the electrical conductance of the skin \cite{boucsein1992electrodermal}, usually performed with one or two sensors attached to some part of the hand or foot \cite{Nourbakhsh2012}. Skin conductivity varies with changes in skin moisture level (sweating) which can reveal changes in sympathetic nervous system related to arousal \cite{Lang1990,Gainesville2008}. Changes in GSR are related to the presence of emotions such as stress or surprise \cite{Gainesville2008}.
		
		\textbf{ECG}\label{ECG_modality}: Electrocardiogram is a recording of the electrical activity of the heart. It is detected by electrodes attached to the skin surface, which pick up electrical impulses generated by the polarization and depolarization of cardiac tissue.
		ECG can reveal changes of the autonomous nervous system related to affective experiences and stress \cite{Bong2012}.
		
		In previous databases, neuro-physiological signals have been recorded using laboratory equipment (e.g. Biosemi ActiveTwo) which is expensive and limits the mobility of the participants. 
		In this database, the neuro-physiological signals have been recorded using wearable sensors that allow more freedom given that they use wireless technology. 
		EEG was recorded using the Emotiv EPOC Neuroheadset\footnote{ http://www.emotiv.com/} (14 channel, 128 Hz, 14 bit resolution). EEG channels according to the 10-20 \cite{Ahmad2015wave} system are: AF3, F7, F3, FC5, T7, P7, O1, O2, P8, T8, FC6, F4, F8, AF4. 
		ECG was recorded using the Shimmer 2R\footnote{ http://www.shimmersensing.com/} platform extended with an ECG module board (256 Hz, 12 bit resolution), which uses three electrodes, two of them are placed at the right and left arm crooks and the third one at the internal face of the left ankle as reference. This set-up allows precise identification of heart beats as well as the full ECG QRS complex. 
		GSR signal recorded using the Shimmer 2R platform extended with a GSR module board (128 Hz, 12 bit resolution), with two electrodes placed at the middle phalanges of the left hand's middle and index fingers.

	\subsection{Video Recordings}\label{Video_Modality}
\begin{figure*}[htp]
			\centering
			\setlength{\tabcolsep}{3pt}
			\begin{tabular}{ccccccc}
				
				\includegraphics[height=0.09\paperwidth,keepaspectratio]{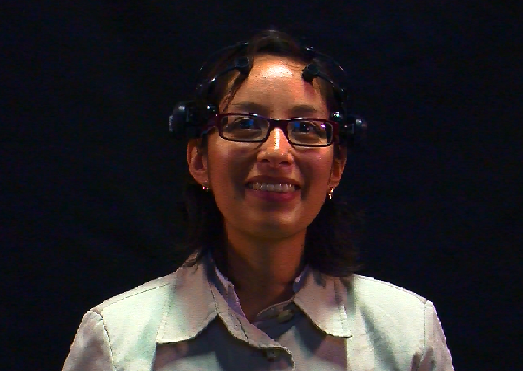} &  \scalebox{-1}[1]{\includegraphics[height=0.09\paperwidth,keepaspectratio]{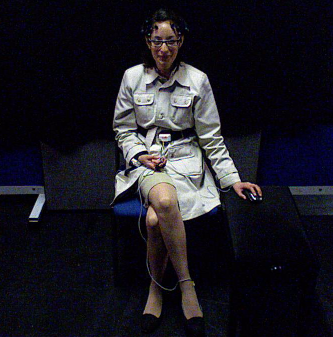}}	&  \scalebox{-1}[1]{\includegraphics[height=0.09\paperwidth,keepaspectratio]{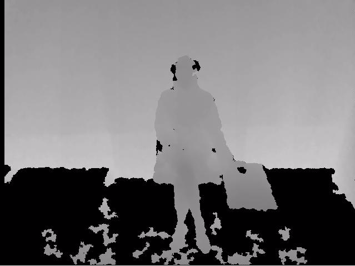}} & & \includegraphics[height=0.09\paperwidth,keepaspectratio]{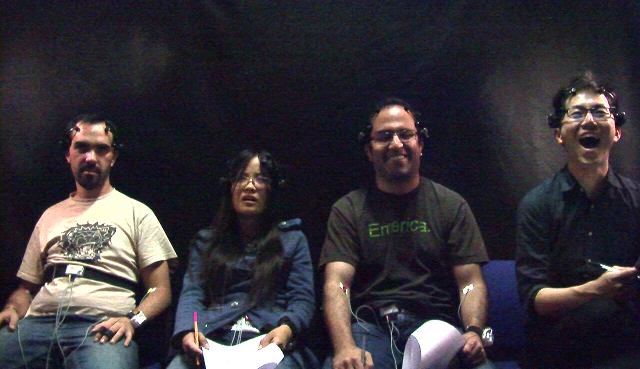} &  \scalebox{-1}[1]{\includegraphics[height=0.09\paperwidth,keepaspectratio]{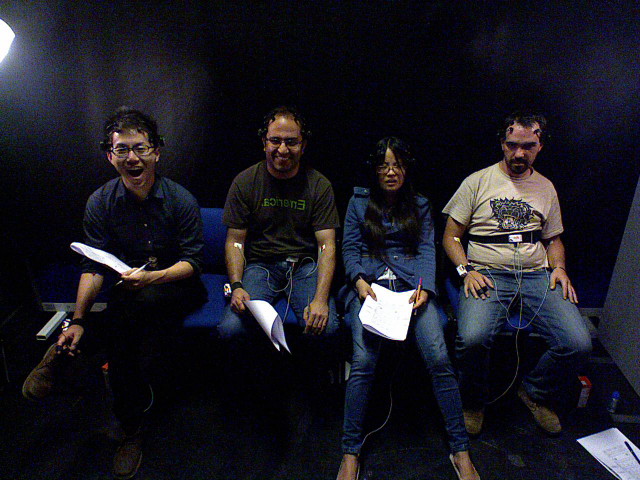}}	&  \scalebox{-1}[1]{\includegraphics[height=0.09\paperwidth,keepaspectratio]{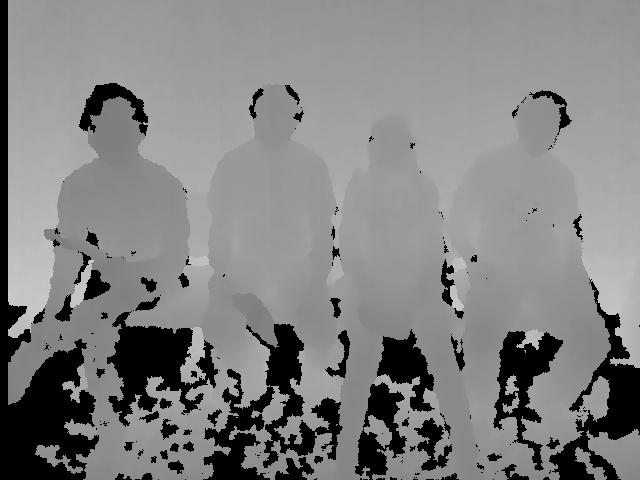}} \\
				(a) & (b) & (c) &  & (d) & (e) & (f)   \\ 
			\end{tabular}
			
			\caption{Participant in experiment conditions during the short videos experiment recorded in (a) Frontal HD video, (b) full body RGB video via Kinect, (c) full body depth video via Kinect; and group of 4 participants during the long videos experiment recorded in (d) frontal HD video, (e) full body RGB video via Kinect and (f) full body depth video via Kinect. }\label{participants_fig}
		\end{figure*}

		Frontal face video was recorded in HD quality using a JVC GY-HM150E camera, positioned just below the screen. Additionally, both RGB and depth full body videos were recorded using a Microsoft's Kinect V1\footnote{ http://developer.microsoft.com/windows/kinect/hardware} placed at the top of the screen. Though this study does not use the visual modality, Mou et al \cite{Mou2016,Mou_2016_CVPR_Workshops} have explored the visual modality on our dataset for prediction of affect, social context and group belonging. A participant during the short videos experiment and a group of participants during the long videos experiment can be observed in Fig.~\ref{participants_fig}.
		
	\subsection{Synchronization and Stimuli Display Platform}\label{Synchronization}
		One PC (Intel Core i7, 3.4 GHz) was used to (i) present the stimuli, (ii) get and synchronize signals, and, in the case of the short videos experiment, (iii) obtain the self-assessment of participants.
		Shimmer sensors were paired to the PC using the bluetooth standard, while the Emotiv headset was paired using a proprietary wireless standard.
		Videos were presented in a 40-inch screen (1280$\times$1024), each of them was displayed preserving the original aspect ratio and covering the highest screen-area possible. The remaining area was filled with black background. Subjects were seated approximately 2 meter from the screen. Stereo speakers were used and the sound volume was set at a relatively loud level, however it was adjusted when necessary.

	
	\subsection{Short Videos Experiment Protocol}\label{protocolshort}
	Recordings were performed in a laboratory environment with controlled illumination. 40 healthy participants (13 female), aged between 21 and 40 (mean age 28.3), took part in the experiment. Prior to the recording session, the participants read and signed a consent form. Then they read a sheet with instructions about the experiment, and an experimenter answered their questions. When the instructions were clear, the participants were led into the experiment room. After that, the experimenter explained the affective scales used in the experiment and how to fill in the self-assessment form (See~\ref{selfassessment}). Next, the sensors were placed and their signals checked with a test recording to assess the quality of the signals. Finally, the experimenter left the room and the recording session began.
	
	The participants performed an initial self-assessment for arousal, valence and dominance, as well as selection of basic emotions (Neutral, Happiness, Sadness, Surprise, Fear, Anger and Disgust) they felt before any stimulus have been shown. Next, 16 videos were presented in a random order in 16 trials, each consisting of:
	(1) A 5 second baseline recording showing a fixation cross.
	(2) The display of a small video.
	(3) Self-assessment of arousal, valence, dominance, liking and familiarity as well as selection of basic emotions (See~\ref{selfassessment}).
	After the 16 trials, the recording session ended. 
	
	\subsection{Long Videos Experiment Protocol}\label{protocollong}
	\changemarker{The participants that took part in the short videos experiment, performed the long videos experiment in either individual or group settings}. In the individual setting, participants performed the experiment alone. In the group setting, participants performed the experiment together with 3 other participants. Only 37 participants took part in the long videos experiment (participants 8, 24 and 28 were not available), 17 of them in individual setting and 20 in group setting (5 groups of 4 people). In order to maximize interactions, groups were formed to include people that knew each other, being either friends, colleagues, or people with similar cultural background \cite{buck1992}. The IDs of participants that were in the individual setting and in each group of the group setting are listed in Table~\ref{GroupMembers}. 
	
	\newcommand{\minitab}[2][l]{\begin{tabular}{#1}#2\end{tabular}} 
	\begin{table}[t]
		\centering
		\fontsize{8}{8}\selectfont
		\renewcommand{\arraystretch}{1.2}
		\caption{\label{GroupMembers} Participant IDs for Individual and Group Settings of the long videos experiment. In the group setting, the IDs order represent the order in which participants were seated, from a front view, from left to right.}
		\vspace{-.3cm}
		\scalebox{0.82}[0.88]{
			\begin{tabular}{|>{\centering\arraybackslash}p{.17\linewidth}|>{\centering\arraybackslash}p{0.18\linewidth}|>{\centering\arraybackslash}p{.22\linewidth}|>{\centering\arraybackslash}p{0.43\linewidth}|}
			\hline
			\hline
			\small
			& Part. ID & & Part. ID \\
			\hline
			Group 1  & 7, 1, 2, 16 & Group 5  & 15, 11, 12, 10 \\ \hline
			Group 2  & 6, 32, 4, 3 & \multirow{3}{\linewidth}{\centering Individual Participants}  & \multirow{3}{\linewidth}{\centering 9, 13, 19, 20, 23, 25, 26, 30, 21,  33, 34, 35, 36, 37, 38, 39, 40} \\ \cline{1-2}
			Group 3  & 29, 5, 27, 21 &   &  \\ \cline{1-2}
			Group 4  & 18, 14, 17, 22 &   &  \\
			\hline
		\end{tabular}
		}
		\vspace{-.3cm}
	\end{table}
	
	During the recording sessions, the participant(s) was(were) led to the recording room. While the different sensors were set up, experimenters explained the differences of the protocol compared to the short videos experiment. Every participant was given a set of self-assessment paper forms (See~\ref{selfassessment}) and a pen, that were used to assess their affective state at the beginning and at the end of each video. Experimenters avoided to mention whether the participants could talk during the experiment, for the interactions to be spontaneous. Once the sensors had been tested, the experimenters left the room and the recording session started.
	
	The experiment consisted of the display of 4 long videos in random order. Videos were shown in two recording sub-sessions, each consisting of: (1) initial self-assessment (45s) of arousal, valence, dominance and selection of basic emotions. (2) the display, in two trials, of two long videos, each followed by (3) self-assessment (45s) of arousal, valence, dominance, liking and familiarity, and selection of basic emotions (See~\ref{selfassessment}). After the first sub-session a break of 15 minutes was given for the participants to rest. During this time they were offered refreshments. After the break, sensors' signals were checked and the second recording sub-session started, after which the experiment ended.
	
	After the long videos experiment, participants were asked to fill in as soon as possible, on-line forms with Personality Traits \cite{McCrae1992} and PANAS \cite{Watson1988} questionnaires (See~\ref{Personality}). Participants took 2 days on average to fill in the forms. Once they filled in all required forms, they were given mugs and university gadgets in return for their participation.

	\subsection{Affective Annotation}\label{annotations}
	Internal annotation (self-assessment) is the process were a subject directly assess its affective state while performing a task \cite{koelstra2010}. It has the advantage of being an easy, and possibly, the most direct way to assess affective states. At the same time, it is an intrusive process, subjects could be unreliable at reporting their emotions or they could hide their real emotions \cite{SoleymaniP12}. External annotation (implicit assessment) is a process that intends to assess a person's affective state without it being actively involved in the process. The assessment is performed by external means such as analyzing the person's behavior and/or its physiological responses \cite{SoleymaniMAHNOB2012}. We have performed both internal and external annotations to assess the participants' affective state. 
	
	\subsubsection{Participant's Affect Self-assessment}\label{selfassessment}
	At the beginning of the recording session of the short videos experiment, and of each of the two recording sub-sessions of the long videos experiment, participants performed a self-assessment of their levels of arousal, valence and dominance, and were asked to select basic emotions that described what they were feeling at the start of each session/sub-session. Then, at the end of each trial, participants performed a self-assessment of the same dimension as the initial self-assessment, and of the liking and familiarity that described what they felt during each video.
	
\begin{figure}[tb]
		\centering
		\includegraphics[width=0.40\paperwidth]{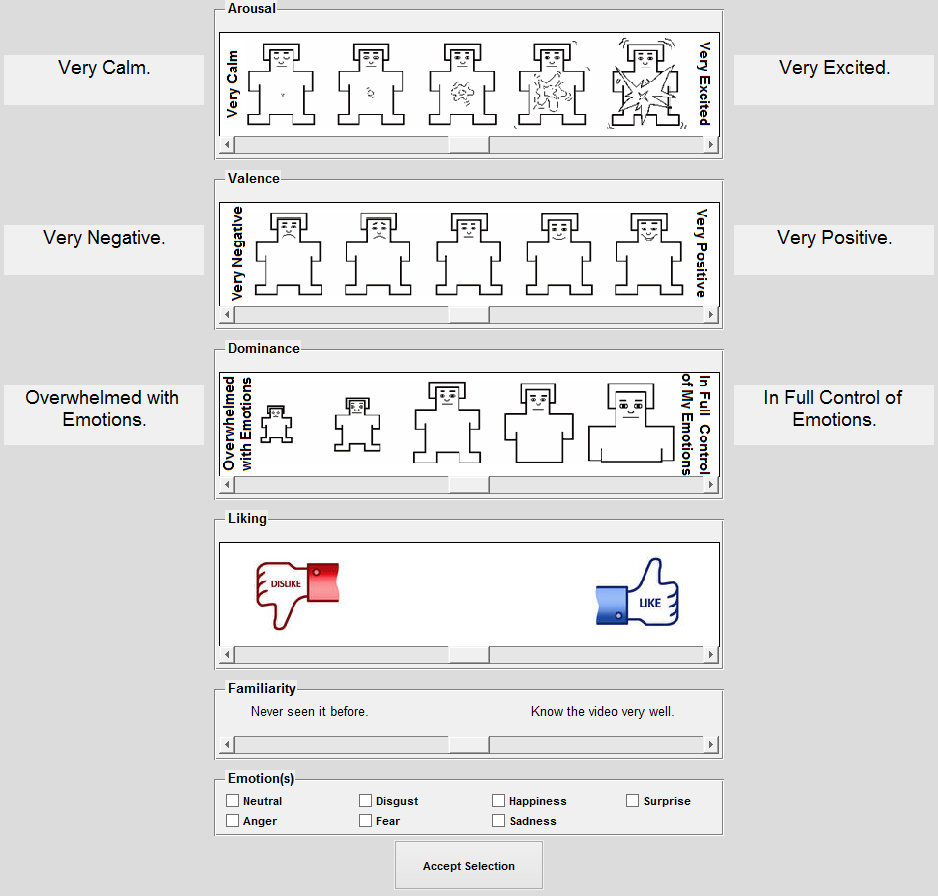}
		\caption{Self-Assessment Form for Assessment of Arousal, Valence, Dominance, Liking, Familiarity and Basic Emotions.}\label{selfAssessmentForm}
	\end{figure}

	The self-assessment form used for the short videos experiment can be seen in Fig.~\ref{selfAssessmentForm}. Self-assessment manikins (SAM) \cite{SAM1995} were used to visualize the scales of valence, arousal and dominance. For the liking scale, thumbs down/thumbs up symbols were used. This inquires the participants' tastes, not feelings. The fifth scale asks the participants to rate their familiarity with the video. Arousal scale ranges from ``very calm'' (1) to ``very excited'' (9). Valence from ``very negative'' (1) to ``very positive'' (9). Dominance from ``overwhelmed with emotions'' (1) to ``in full control of emotions'' (9). The fourth scale ranges from disliking (1) to liking (9) the video. The familiarity scale ranges from ``Never seen it before'' (1) to ``Know the video very well'' (9). Participants moved a continuous slider, placed at the bottom of each scale, to specify their self-assessment level. They were informed they could move the slider anywhere directly below or in-between of the manikins. Finally, participants were asked to select at least one of the basic emotions (Neutral, Disgust, Happiness, Surprise, Anger, Fear and Sadness \cite{Ekman1975utf}), or as many as they felt during the video (a participant can consider a video to be both surprising and sad). 
	
	In the long videos experiment, having a digital form for every participant of the groups was not practical, therefore we opted to use a paper version of the form in Fig.~\ref{selfAssessmentForm} in both individual and group setting recordings, in order to keep consistent the self-assessment between settings. 
	
	In total, for the short videos experiment 17 annotations were obtained from each participant (1 at the beginning of the experiment and 1 after each of the 16 short videos), and 6 annotations in the case of the long videos experiment (1 at beginning of the first recording sub-session, 1 after each of the two long videos of the first recording sub-session, 1 at the beginning of the second recording sub-session just after the 15 minute break and 1 after each of the two long videos of the second recording sub-session). It is important to note that this annotation gives information related only to the participants' initial and final affective states, not for specific instants during the videos.
	
	\subsubsection{External Affect Annotation}\label{Annotation}
	In order to study the temporal evolution of affect, the frontal videos of each participant recorded during the display of the stimuli of both experiments were off-line annotated on the valance and arousal dimensions as follows. 
	
	First, the videos of a given participant recorded during the display of each of the $20$ stimuli videos ($16$ short and $4$ long), were manually cropped in order to show only a squared region around the face, covering from the top of the head to the start of the shoulders. Then each of the participants' face videos were split into $20$ second clips. For this, the first $20$ seconds of each video, including 5 seconds prior to the presentation of the stimuli, were extracted as first clip, then, starting from the $5$s of the video (instant in which the stimuli started), $n=\lfloor{(D)/(20s)}\rfloor$ non overlapping segments of $20$s were extracted, with $D$ being the duration of the stimuli video in seconds. Finally, the last $20$ seconds of the video were extracted as final clip. For every participant, \{6, 7, 5, 6, 4, 5, 8, 5, 7, 5, 9, 5, 5, 4, 6, 7, 72, 58, 72 and 44\} clips were obtained respectively from videos \{4, 5, 9, 10, 13, 18, 19, 20, 23, 30, 31, 34, 36, 58, 80, 138, N1, P1, B1 and U1\}, totaling 340 clips per participant, 94 corresponding to the short and 246 to the long videos experiment. 
	
	Three annotators rated on the valence and arousal scales the clips of all the participants (340 clips $\times$ 37 participants $=$ 12580 clips). Both scales were continuous and ranged from $-1$ (low valence/arousal) to $1$ (high valence/arousal). The 340 clips of a given participant, were annotated in the same random order by each annotator, however, the order of the clips was different for each participant. Since samples of both experiments were randomly shown to the annotators, labels of the two experiments are directly comparable. The pipeline of the annotation consisted of the display of a randomly selected clip followed by the annotation performed by the annotator, first, of valence and then of arousal. This process was repeated until all clips were annotated. 
	
	\subsection{Personality and Mood Assessment}\label{Personality}
	The Big-Five personality traits were measured with an on-line form of the big-five marker scale questionnaire \cite{Perugini2002}, in which, for each personality trait, using the basic question ``I see myself as a person:", ten descriptive adjectives are rated with a 7-point-likert-scale \cite{likert1932technique} and a mean is calculated. 
	
	Mood was assessed on the positive affect (PA) and negative affect (NA) schedules (PANAS) \cite{Watson1999} model, using an on-line form of the general PANAS questionnaire \cite{Watson1999} which consists of two 10 questions sets, each to access the PA and NA respectively. Participants rated their general feelings in a 5-point intensity scale using questions like ``Do you feel in general...?" (e.g. active, afraid See~\cite{Watson1999}). 
	PANAS is calculated by summing the ratings of all 10 questions for PA and NA respectively, resulting in values between 10 and 50. 
	
\begin{figure}[b]
		\flushleft
		\includegraphics[width=\linewidth]{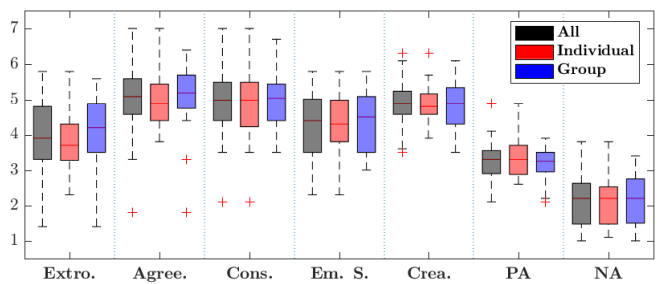}
		\caption{\label{Distr} Distribution of the Big-Five Personality Traits (Extraversion, Agreeableness, Conscientiousness, Emotional Stability and Openness) and Positive Affect and Negative Affect Schedules (PA and NA) for (i)~All, (ii)~Individual setting, and (iii)~Group setting participants of the Long Videos Experiments. PA and NA are scaled by a $0.1$ factor.}
	\end{figure}

	The distribution of the Big-Five personality traits, PA and NA, over (i) the 37 participants that took part in the long videos experiment, (ii) the 17 participants of the individual setting, and (iii) the 20 participants of the group setting, are presented in Figure~\ref{Distr}. Note that PA and NA scores have been scaled by a $0.1$ factor. The difference of distribution of ratings, for each of the seven dimensions of personality and PANAS, between the participants of individual and group settings, is not significant ($p>0.1$ according to a two sample t-test for every dimension). 
	
	\def\esc{0.9}

	\section{Data Analysis}\label{Analysis}
		In this section, we present a detailed analysis of the data gathered in both experiments.
		
		\subsection{Self-Assessment vs External Annotation}\label{Annotation_vs_annotation}
			The external annotations were validated by assessing the inter-annotator agreement. For this, the annotations corresponding to each participant performed by every annotator were mapped to the $[0,1]$ range, where $0$ corresponds to low and $1$ to high valence(arousal), then the Cronbach's $\alpha$ \cite{Cronbach1951} statistic among annotators, commonly used for agreement assessment on continuous scales \cite{McKeown2012}, was calculated. 
			Mean Cronbach's $\alpha$s over all participants of $0.98$ for valence and $0.96$ for arousal were obtained, which indicates a very strong inter annotator reliability for both dimensions. 

			With the objective to test at what degree, the affective state of participants assessed through self-assessment, is represented by the external annotations, a comparison between the self-assessment and external annotations of valence and arousal, for the short videos experiment, was performed. For each participant, the Spearman correlation coefficient as well as the \textit{p}-value for the positive correlation test were calculated between the self-assessment scores of each video and the mean external annotation over all the annotators and segments of each video. Assuming independence, the resulting \textit{p}-values were combined to one \textit{p}-value using Fisher's method \cite{Loughin2004}. For valence, the mean correlation over all participants is $0.44(p<.05)$, and $0.15(p<.05)$ for arousal. These correlations are statistically significant which indicates that the external annotation is a good predictor of the affective state of participants, though for the arousal dimension the correlation is low which shows that it is easier to externally assess valence than arousal. 
			
\begin{figure}[tp]
				\setlength{\tabcolsep}{1pt}
				\centering
				\begin{tabular}{cccc}
					\rotatebox[origin=c]{90}{Arousal} & \raisebox{-.5\height}{\includegraphics[width=0.45\linewidth]{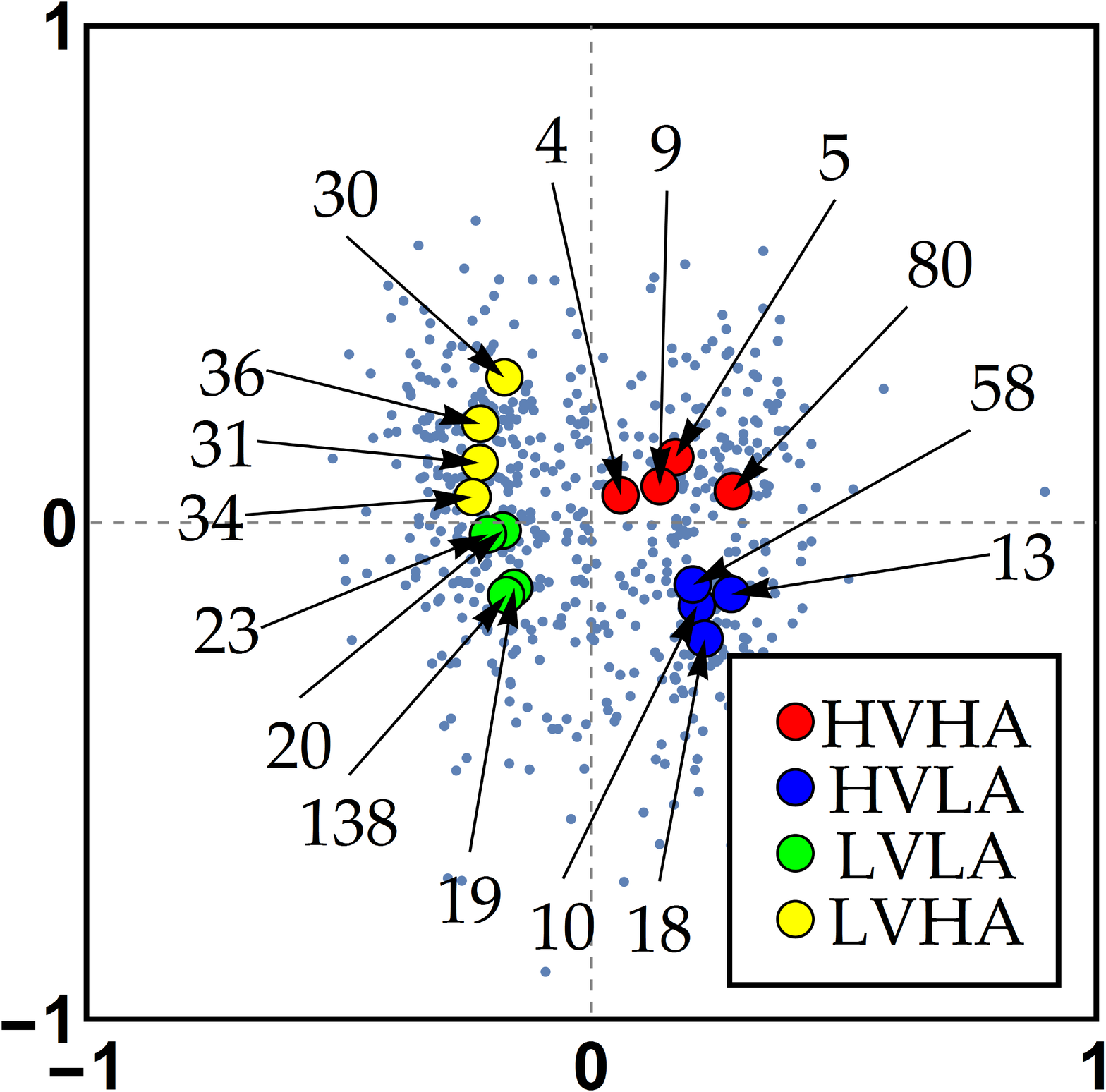}}	& \rotatebox[origin=c]{90}{Arousal} & \raisebox{-.5\height}{\includegraphics[width=0.45\linewidth]{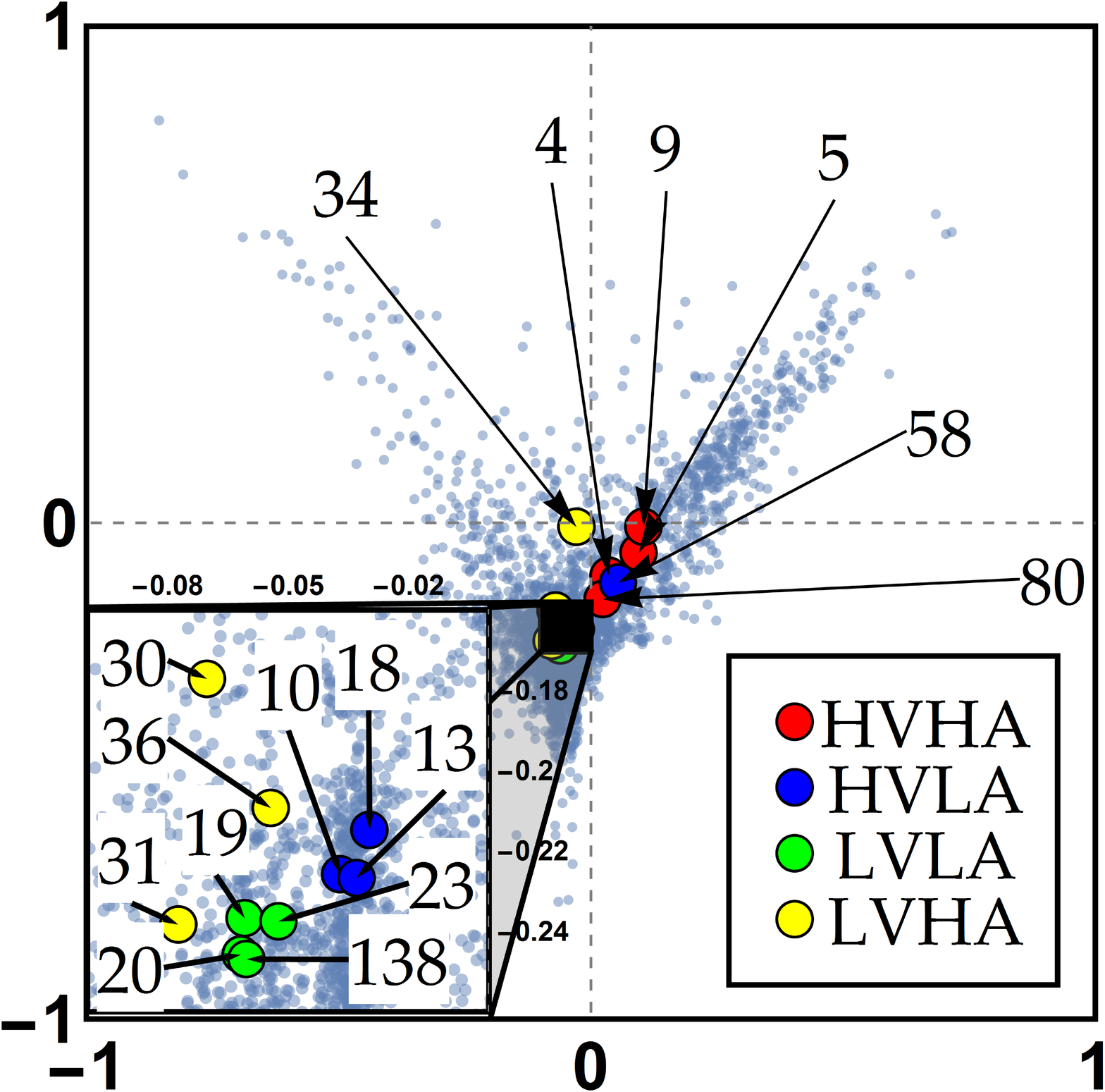}}\\
					& Valence	& & Valence\\
					&(a)	& & (b)
				\end{tabular}
				\caption{Distribution of ratings of Valence vs Arousal, for (a) participants' self-assessment of the $16$ short videos experiment, and (b) mean external annotations over all annotators for $94$ twenty-second segments of the videos of the short videos experiment. Small circles indicate the mean scores over all participants for each of the videos (video ID indicated through arrows). Circles are color coded according to the expected affective response (See Table~\ref{VideoList}). H, L, V and A, refer to high, low, valence and arousal. }\label{externalAnnFig}
			\end{figure}

			In Figure~\ref{externalAnnFig}(a), the distribution of the self-assessment of valence and arousal of all participants for the short videos experiments (16 samples per participant) can be observed. Annotations of each participant have been mapped to the $[-1,1]$ range. The graph includes circles representing the mean scores, over all participants, of each video. It can be observed that in general valence elicitation worked better than arousal, showing a well defined separation between low and high valence stimuli. Even though the separation of arousal is not as prominent, still there is a difference between low and high arousal stimuli. Figure~\ref{externalAnnFig}(b) shows the distribution of the external annotations of valence and arousal over the 16 videos of the short videos experiment (94 samples by participant). The mean scores, over all the 20-second clips of each video and all the participants are marked with circles. It can be observed that the data shows a V-shape relating valence and arousal, which is a result of the difficulty of eliciting high-levels of arousal with neutral valence, and high/low levels of valence with low arousal. It can also be observed that in general participants showed the expected affective states (e.g. participants showed higher valence(arousal) with high valence(arousal) content in comparison to low valence(arousal) content), though the difference is not as clear as in self-assessment (Fig.~\ref{externalAnnFig}(a)).

		\subsection{Analysis of Valence and Arousal for Individual and Group Settings}\label{Correlations}
		
			The external annotations of both experiments have been analyzed to test if valence and arousal, expressed by the participants, differed depending on the social context. Two sets of participants were considered. The first set (individual set) corresponds to the 17 participants that took part in the long videos experiment in individual setting, and the second set (group set) corresponds to the 20 participants took part in group setting. 
			
\begin{figure*}[htp]
				\centering
				\scalebox{0.88}[0.88]{
					\begin{tabular}{cccc}
						\rotatebox[origin=c]{90}{\scalebox{0.7}[0.7]{LV $\longleftarrow$ Valence $\longrightarrow$ HV}}&\raisebox{-.5\height}{\includegraphics[height=0.18\linewidth]{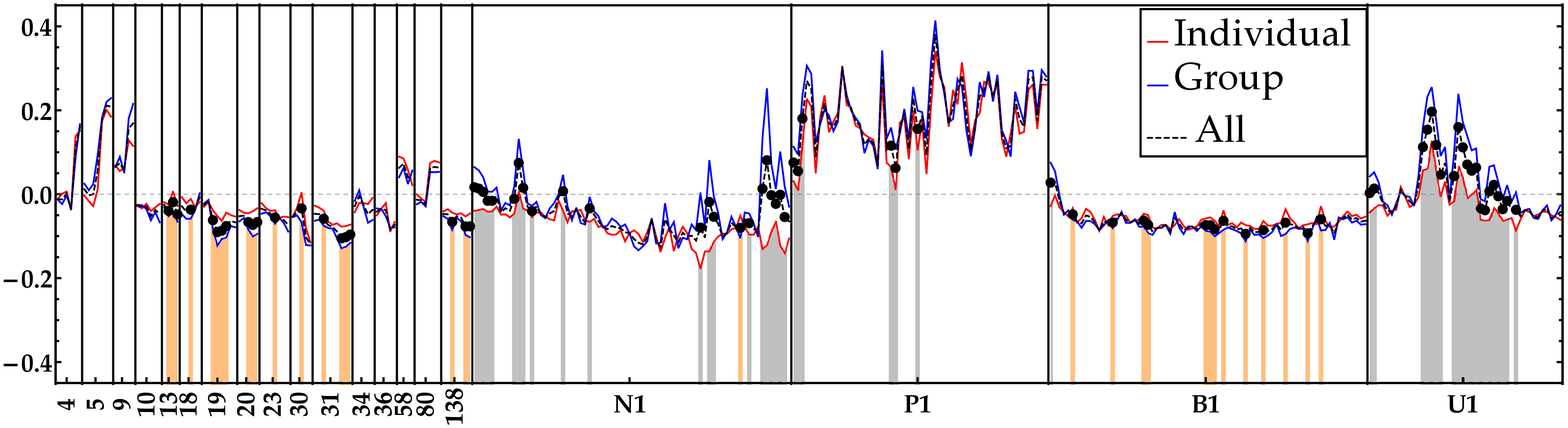}}	& \raisebox{-.5\height}{\includegraphics[height=0.18\linewidth]{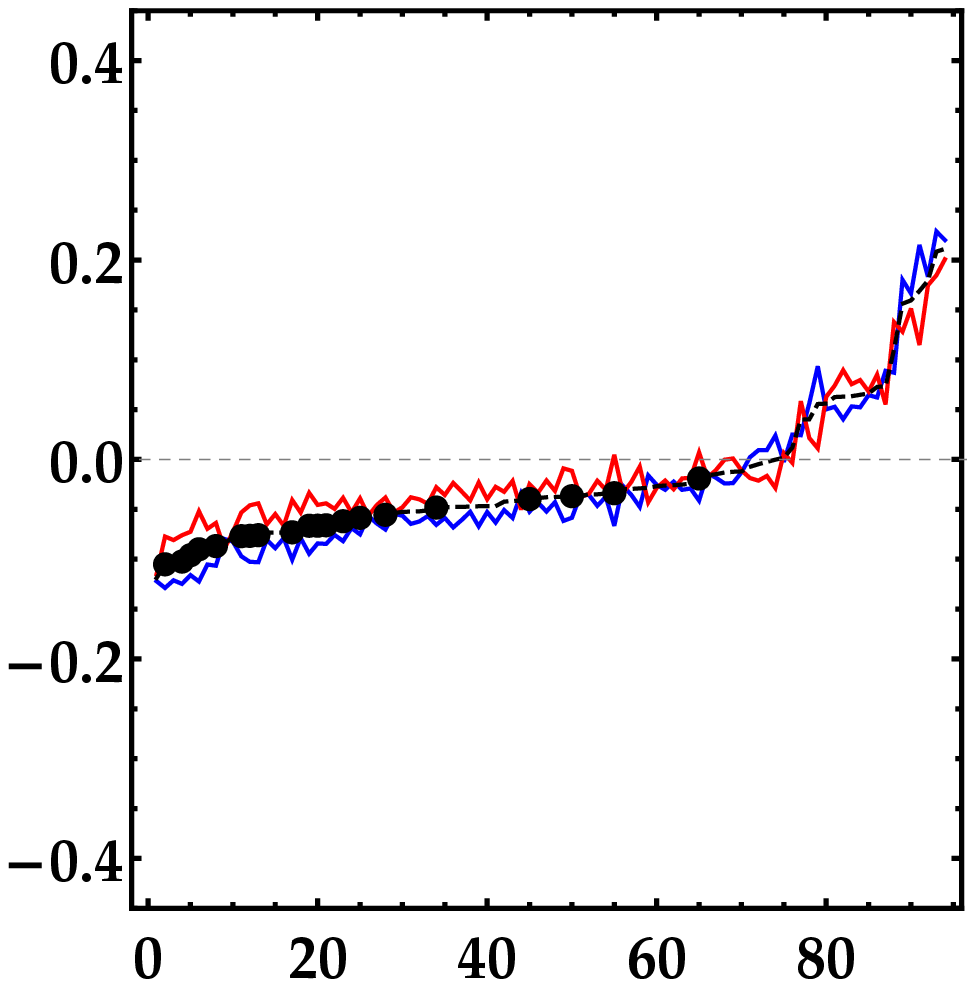}}	& \raisebox{-.5\height}{\includegraphics[height=0.18\linewidth]{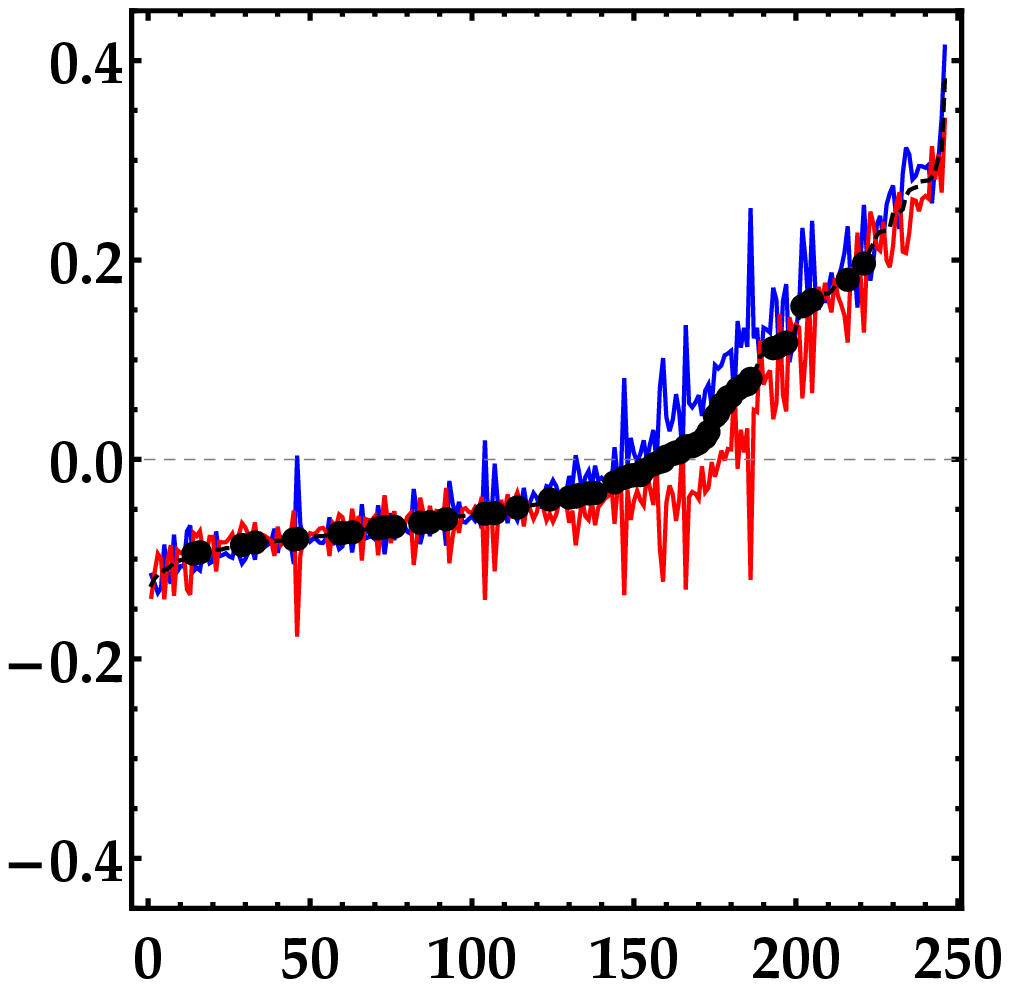}}\\
						&(a)	& (b) & (c) \\
						\rotatebox[origin=c]{90}{\scalebox{0.7}[0.7]{LA $\longleftarrow$ Arousal $\longrightarrow$ HA}}&\raisebox{-.5\height}{\includegraphics[height=0.18\linewidth]{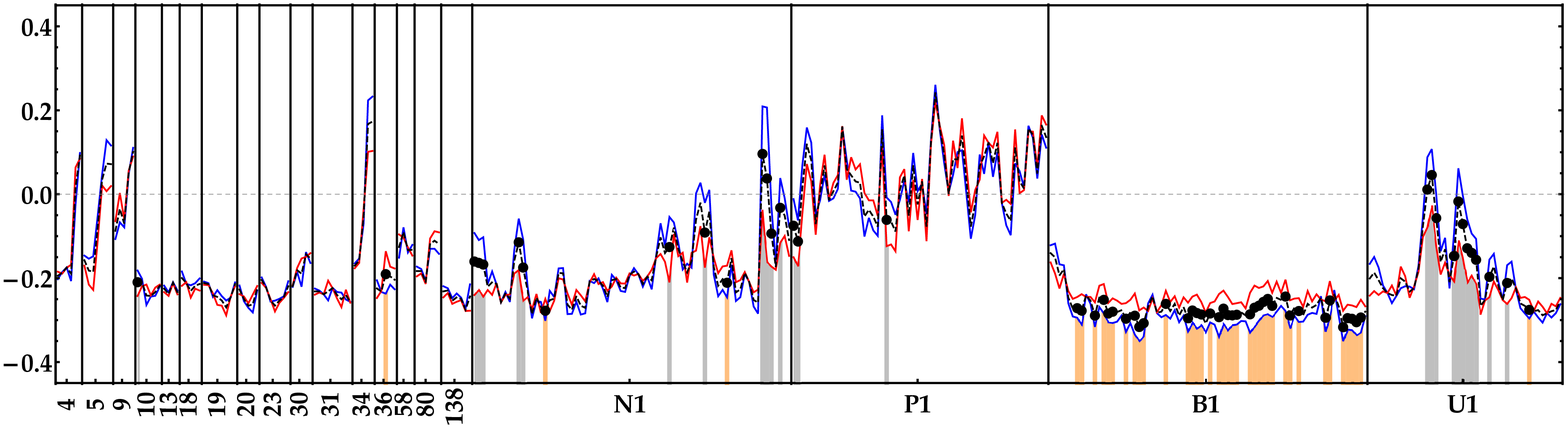}}	& \raisebox{-.5\height}{\includegraphics[height=0.18\linewidth]{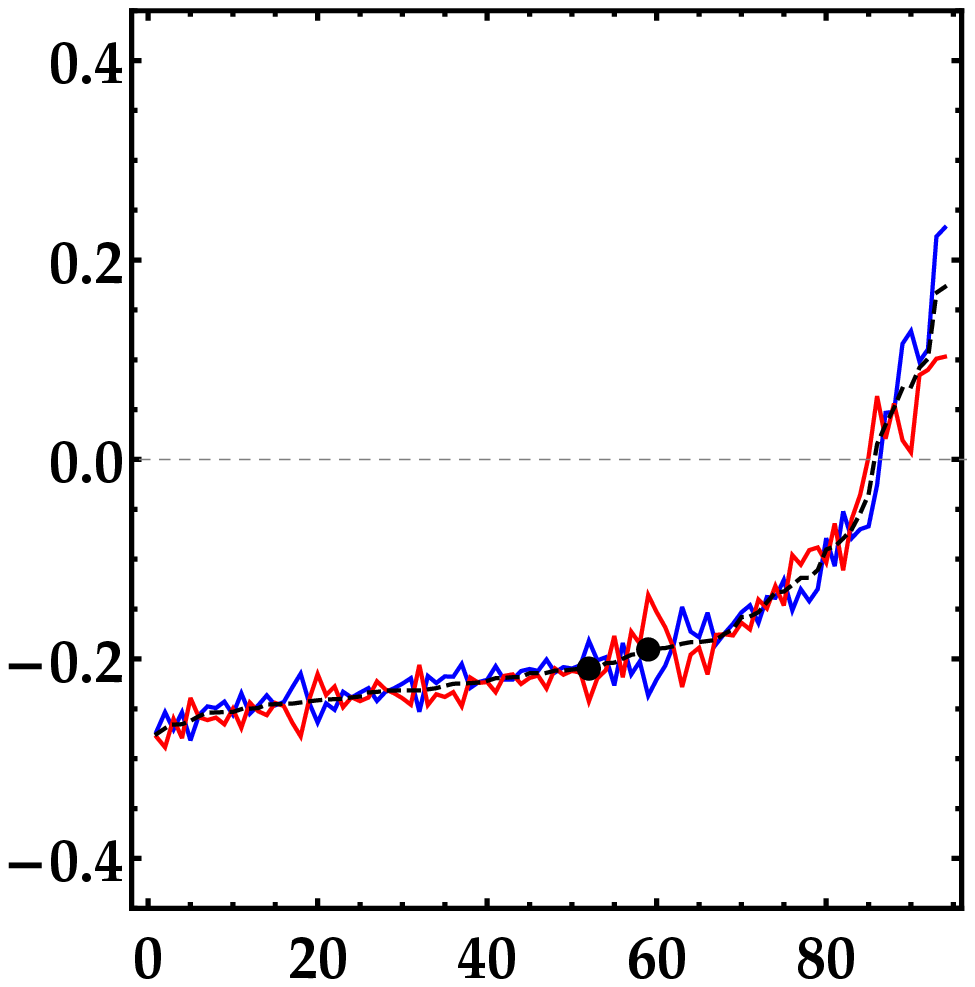}}	& \raisebox{-.5\height}{\includegraphics[height=0.18\linewidth]{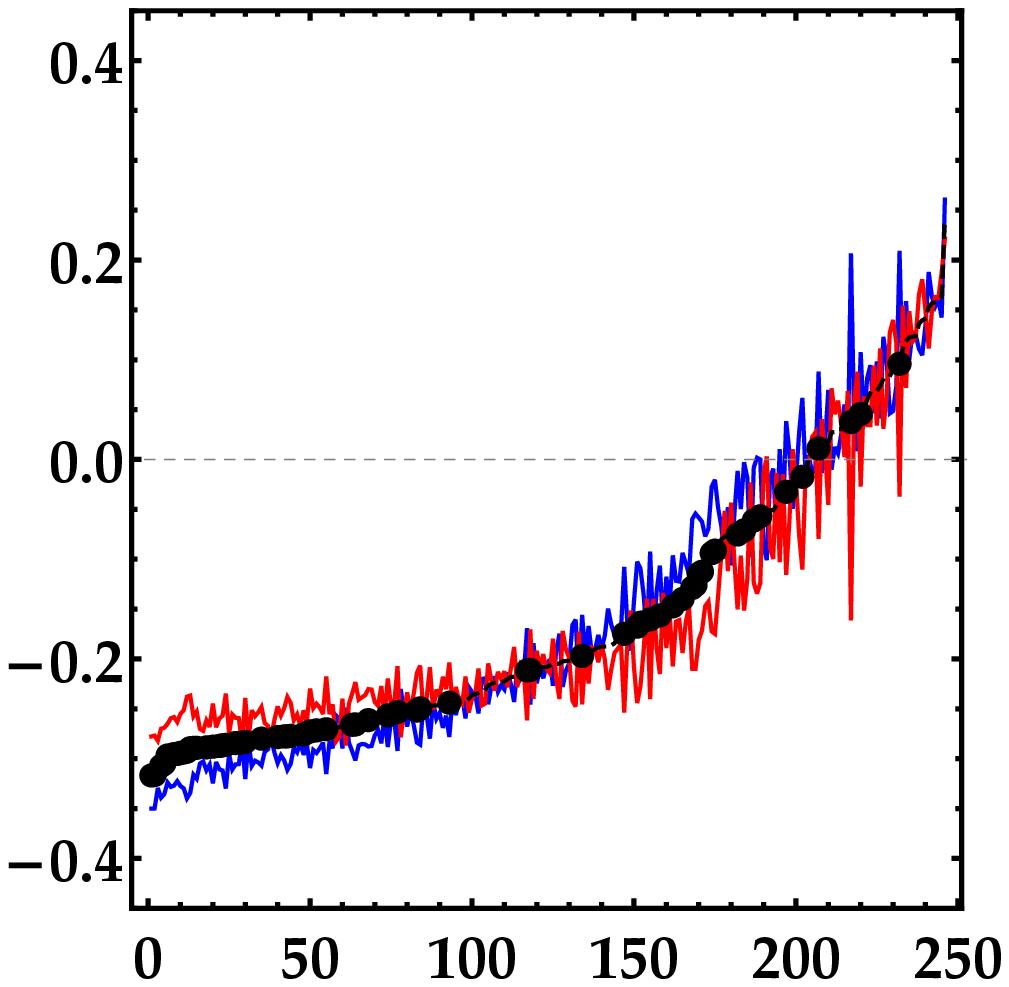}}\\
						&(d)	& (e) & (f)
					\end{tabular}
				}
				\caption{Mean external annotations of Valence ($V$, upper graphs (a), (b) and (c)) and Arousal ($A$, lower graphs (d), (e), and (f)), over individual participants (red curve), group participants (blue curve) and all participants (dashed black curve), for the videos of ((a) and (d)) both short and long videos experiments (340 segments), ((b) and (e)) the short videos experiment (94 segments), and ((c) and (f)) the long videos experiment (246 segments). Clips where the distribution of scores of individual participants is significantly different than the one of group participants ($p<0.05$ according to a two sample t-test), are marked with black points. In the case of (a) and (d), video IDs are indicated in the captions. Clips where the distribution of scores of individual participants is significantly higher than the one of group participants ($p<0.05$), are highlighted in orange. Clips where the distribution of scores of group participants is significantly higher than the one of individual participants are highlighted in gray. In the case of (b), (c), (d) and (f) the horizontal axis represent the number of clips. Origin of valence and arousal (horizontal axis at ($V=0$) and ($A=0$)) divides the scale into high-valence (HV: $V>0$) and low-valence (LV: $V<0$), and into high-arousal (HA: $A>0$) and low-arousal (LA: $A<0$).
				}\label{groupsVsIndividual}
			\end{figure*}

			In Fig.~\ref{groupsVsIndividual}, the differences in annotations of valence and arousal for the individual set in comparison with the group set for both short and long videos experiments are shown. Fig.~\ref{groupsVsIndividual}(a) and (d) show the mean valence and arousal annotations for (i) the individual set (red curve), (ii) the group set (blue curve), and (iii) all participants (black dashed curve), for each of the 340 20s clips. The clips are shown by the video they are part of and ordered according their appearance in the video. In the figure, clips where the difference in the distribution of scores for the group set are significantly lower or higher ($p<0.05$ according to a two sample t-test) with respect to the one of the individual set are marked with black points and have been shadowed (orange for group scores $<$ individual scores and gray for group scores $>$ individual scores). Fig.~\ref{groupsVsIndividual}(b) and (e), show the mean annotations of valence and arousal, for the same sets of participants, of the clips of the short videos experiment, whereas Fig.~\ref{groupsVsIndividual}(c) and (f) present the mean annotations for the clips of the long videos experiment. In the (b), (c), (e) and (f) graphs, samples are ordered according to the mean score over all participants (dashed black curve). The clips for which the difference between the distribution of scores from individual and group sets is significant ($p<0.05$ according to a two sample t-test) are marked with black points.
			
		From Fig.~\ref{groupsVsIndividual}(a) and (d) it can be observed that both the high and low areas of the valence and arousal dimensions are covered between all the videos. Comparing the graphs of the short videos experiment (Fig.~\ref{groupsVsIndividual}(b) and (e)) with the ones of the long videos experiment (Fig.~\ref{groupsVsIndividual}(c) and (f)), it can be observed that in the short videos experiment, where all participants were alone, $21.3\%$ of the clips present significant differences in valence between group and individual participants, and they are concentrated in the low valence region, and $2.1\%$ of the clips present significant differences in arousal. In the long videos experiment, where some participants were in groups,  $25.6\%$ of the clips present significant difference of valence between groups and individuals. It is important to note that $48\%$ the clips with significant differences appear in the high valence region (mean valence $>0$). For arousal, $26.4\%$ of the clips present significant differences between groups and individuals. In Fig.~\ref{groupsVsIndividual}(f), where it is observed that in the long videos experiment, group participants showed lower levels of arousal for low arousal clips as well as higher levels of arousal for high arousal clips than individuals.  
			 
		The Spearman correlation coefficient $\rho$ and the p-value were calculated between the social context label and the mean external annotations for valence and arousal, for the clips of the long videos experiments. The social context label was considered 0 if the participant was in individual setting and 1 if it was in group setting. Significant positive correlation ($\rho=0.37$, $p<0.05$) was found between the social context and the mean valence. This significant correlation implies that, in the long videos experiment, participants in group setting showed higher valence than the ones in individual setting. Significant correlation was not found between social context and arousal scores ($p>0.05$), which suggest that social context does not have a common effect in the arousal expressed by the participants for all clips.

		Fig.~\ref{groupsVsIndividual} (c) and (f) show that the scores for clips with low levels of valence(arousal), present a different behavior than the ones with high levels. Therefore, analyses have been independently performed for the low and high valence(arousal) clips of the long videos experiment. For each of the two dimensions (valence and arousal), the clips were sorted based on their score in increasing order, then half of the clips with the lower scores were classified as low class (e.g. low valence) and the other half as high class (e.g. high valence). A two sample t-test of the mean scores of valence(arousal) were performed between the individual and group settings for the clips of low and high classes of valence(arousal). Significant difference was found between individual and group settings for the high valence ($p<0.001$), low arousal ($p<0.001$) and high arousal clips ($p<0.05$), but not for low valence clips ($p=0.90$). Therefore, social context has an important effect on the valence and arousal expressed by the participants.


		\subsection{Affect, Personality, Mood and Social Context 
		}\label{Relations}
		
			In Table \ref{Corr1}, the Spearman inter-correlations observed between the dimensions of personality, PANAS and social context in the long videos experiment are shown. It also shows the inter-correlations that those dimensions have with the mean external annotations of valence and arousal, of the clips of the short and long videos experiments. 
			
			\begin{table}[t]
				\centering
				\fontsize{7}{8}\selectfont
				\renewcommand{\arraystretch}{1.2}
				\caption{\label{Corr1} Inter-correlation Between the Dimensions of Personality, PANAS, Social Context in the Long Videos Experiment, and By-participant Mean External Annotations for Valence and Arousal of Short Videos and Long Videos. Significant correlations ($p<0.05$) are in bold. Ag. Co. E. S., Op. and S. C. refer to Agreeableness, Conscientiousness, Emotional Stability, Openness and Social Context  respectively.}
				\scalebox{0.88}[0.9]{
					\setlength{\tabcolsep}{2pt}
					\begin{tabular}{|l|c|c|c|c|c|c|c|c|c|c|c|}
						\hline
						\hline
						\multirow{2}{*}{\centering Dims.} & \multirow{2}{*}{\centering Ag.} & \multirow{2}{*}{\centering Co.} & \multirow{2}{*}{\centering E. S.} & \multirow{2}{*}{\centering Op.} & \multirow{2}{*}{\centering PA} & \multirow{2}{*}{\centering NA} & \multirow{2}{*}{\centering S. C.} & \multicolumn{2}{c|}{Valence} & \multicolumn{2}{c|}{Arousal} \\
						\cline{9-12} & & & & & & & & Short & Long  & Short & Long\\
						\hline
						Ex. & \textbf{0.44*} & 0.09 & 0.21 & 0.13 & 0.32 & \textbf{-0.48*} & 0.20 & -0.01 & 0.02 & 0.05 & 0.18 \\
						\hline
						Ag. & - & \textbf{0.34*} & 0.14 & 0.24 & \textbf{0.43*} & \textbf{-0.41*} & 0.18 & -0.21 & 0.00  & 0.13 & 0.21 \\
						\hline
						Co. & - & - & \textbf{0.35*} & -0.01 & 0.26 & -0.26 & 0.07 & -0.12 & 0.14  & 0.13 & 0.19 \\
						\hline
						E. S. & - & - & - & 0.24 & -0.12 & \textbf{-0.64*} & 0.03 & 0.21 & 0.11  & -0.18 & -0.15\\
						\hline
						Op.& - & - & - & - & 0.20 & \textbf{-0.35*} & -0.04 & 0.23 & 0.13 & 0.06 & 0.02 \\
						\hline
						PA & - & - & - & - & - & -0.06 & -0.03 & -0.03 & 0.16  & 0.30 & \textbf{0.61*}\\
						\hline
						NA & - & - & - & - & - & - & -0.01 & -0.28 & -0.02 & -0.12  & 0.04 \\
						\hline					
					\end{tabular}
				}
			
			\vspace{-.3cm}
			\end{table}
			
			For personality and PANAS, positive significant correlations ($p<0.05$) were obtained between extraversion and agreeableness, agreeableness and both conscientiousness and PA, and conscientiousness and emotional stability. NA is negatively correlated to all personality and PA dimensions. For social context, significant differences in personality and PANAS distribution between individual and group participants were not obtained, which imply that the group and individual participants have similar distribution of personalities (e.g. individual and group participants have similar levels of extraversion). In general, correlations between personality and PANAS with respect to valence and arousal were not significant, which implies that personality and mood do not necessarily affect the levels of valence and arousal expressed by the participants, with the exception of PA which showed significant positive correlation ($0.61$) with respect to arousal of the long videos, which indicates that high-PA participants showed higher levels of arousal (they showed more active emotions) than low-PA participants.



\section{Affect, Personality and PANAS Recognition from Neuro-Physiological Signals}\label{Evaluation}
	In this section, our baseline methods and results for prediction of affect (valence and arousal), personality, PANAS and social context using neuro-physiological signals are presented. First, the features extracted from the used modalities are described. Next, our method for single modality and fusion of modalities for single-trial classification of affect is presented. Then, our method for single-trial classification of personality traits, PANAS and social context, using single modalities and different schemes for fusion of modalities is presented. Finally, our results are presented and discussed.

	\subsection{EEG, ECG and GSR Features}\label{Phys_Features}
		The neuro-physiological modalities of EEG, ECG and GSR were used to record the participants' implicit responses to affective content. Below, the extracted features from the employed modalities are described. All the features were calculated using the signals recorded during each of the 340 twenty-second clips described in section \ref{Annotation}. Different to other studies that use the concatenation of ECG and GSR as one modality, we study each of them independently to account for the contribution of each one to the recognition task. The summary of features is listed in Table~\ref{PhysFeat}.
		
		EEG: Following \cite{koelstra2012deap}, power spectral density (PSD) features were extracted from the EEG signals. For this, the EEG data was processed using the sampling frequency of 128 Hz. The signals were average-referenced and high-pass filtered with a 2 Hz cut-off frequency. Eye artefacts were removed with a blind source separation technique \cite{gomez2010blind}. By employing the Welch method with windows of 128 samples ($1.0s$), PSDs, between 3 and 47 Hz, of the signals of every clip were calculated for each of the 14 EEG channels. The obtained PSDs were then averaged over the frequency bands of theta (3-7 Hz), slow alpha (8-10 Hz), alpha (8-13 Hz), beta (14-29 Hz) and gamma (30-47 Hz), and their logarithms were obtained as features. Additionally, the spectral power asymmetry between the 7 pairs of symmetrical electrodes, in the five bands, was calculated.  105 PSD features were obtained (14 channel * 5 bands and 7 symmetrical channels * 5 bands) for every sample (See Table~\ref{PhysFeat}).
		
		ECG: Following \cite{Kim2008}, the heart beats were accurately localized in ECG signals (R-peaks) to calculate the inter beat intervals (IBI). Using IBI values, the heart rate (HR) and heart rate variability (HRV) time series were calculated. Following \cite{SoleymaniMAHNOB2012} and \cite{Kim2008} 77 features were extracted (See Table~\ref{PhysFeat}).
		 
		GSR: Following the method of Kim \cite{Kim2008}, the skin conductance (SC) was calculated from the GSR and then the SC signal was normalized. The normalized signal was low-pass filtered with 0.2 Hz and 0.08 Hz cut-off frequencies to get the low pass (LP) and very low pass (VLP) signals, respectively. Then, the filtered signals were de-trended  by removing the continuous piecewise linear trend in the two signals. 31 GSR features employed in \cite{koelstra2012deap, SoleymaniMAHNOB2012} were calculated  (See Table~\ref{PhysFeat}).

		\begin{table}[t]
			\centering
			\fontsize{8}{8}\selectfont
			\renewcommand{\arraystretch}{1.2}
			\caption{\label{PhysFeat}Extracted Affective Features for each Modality (feature dimension stated in parenthesis). Computed statistics are: mean, standard deviation (std), skewness, kurtosis of the raw feature over time and \% of times the feature value is above/below mean$\pm$std.}
			\vspace{-.3cm}
			\scalebox{0.88}[0.9]{
				\begin{tabular}{|p{.17\linewidth}|p{0.85\linewidth}|}
				\hline
				\hline
				\small
				\centering{\textbf{Modality}} &  \textbf{Extracted features}\\
				\hline
				\centering{\textbf{EEG (105)}} & 	5 bands (theta, slow alpha, alpha, beta and gamma) PSD for each electrode. The spectral power asymmetry between 7 pairs of electrodes in the five bands.	\\  
				\hline
				\centering{\textbf{ECG (77)}} & Root mean square of the mean squared of IBIs, mean IBI, 60 spectral power in the bands from [0-6] Hz component of the ECG signal, low frequency [0.01,0.08]Hz, medium frequency [0.08,0.15] and hight frequency [0.15,0.5] Hz components of HRV spectral power, HR and HRV stats.  \\  
				\hline
				\centering{\textbf{GSR (31)}} & Mean skin resistance and mean of derivative, mean differential for negative values only (mean decrease rate during decay time), proportion of negative derivative samples, number of local minima in the GSR signal, average rising time of the GSR signal, spectral power in the [0-2.4] Hz band, zero crossing rate of skin conductance slow response (SCSR) [0-0.2] Hz, zero crossing rate of skin conductance very slow response (SCVSR) [0-0.08] Hz, mean SCSR and SCVSR peak magnitude. \\  
				\hline
			\end{tabular}
		}
			\vspace{-.3cm}
		\end{table}

	\subsection{ Single Trial Classification of Affect in Short and Long Videos}\label{Facial_Landmarks}
	
		For single trial affect (valence and arousal) classification, the features of every modality for each recording session were mapped to the $[-1,1]$ range in order to avoid the baseline differences that are natural to different recording sessions. This was done for every participant, considering each of the 4 long videos as a recording session and the recordings of the 16 videos of the short videos experiment as a fifth session. For each of the modalities (EEG, ECG and GSR), three scenarios were tested. The first one considers to train and test the system only with the samples of the short videos experiment (94 samples by participant). The second considers only the samples of the long videos experiment (246 samples by participant). The third one considers the combination of the samples of all the videos of both experiments (340 samples by participant), giving in total 9 recognition tasks for every affect dimension. 
		
		Leave-one-participant-out cross validation was used, in which, in order to predict each affect dimension $j$ label, for each participant $i$ a Gaussian ($G$) Na\"{i}ve Bayes (NB) classifier is trained. A NB $G$ assumes independence of the features and is given by: 
		
		\scalebox{0.97}{$G(f_1,...,f_n)=\argmax_{c} p(C=c)\prod_{i=1}^{n}p(F_i=f_i|C=c)$}
		
		where $F$ is the set of features and $C$ the classes. $p(F_i=f_i|C=c)$ is estimated by assuming Gaussian distributions of the features and modeling these from the training set. In each step of the cross validation, from the $N$ available participants, the samples of one participant are used as the test set and the samples of the remaining $N-1$ participants are used as the training set. 
		
		For feature selection, Fisher's linear discriminant $J$ \cite{Song2010} defined as $J(f)=\dfrac{|\mu_1-\mu_0|}{\sigma_1^2+\sigma_0^2}$ is calculated for each feature from the training samples. Features are then sorted in decreasing order according to their $J$ value and with a second 10-fold cross-validation over the training set, the optimal $[1:h]$ most discriminative features are selected. Then, the classifier is trained over all the samples of the training set using the selected features, then it is tested in the test set.
		
		For each of the three scenarios (short, long and all videos), feature level fusion of modalities has also been explored, in which, previous to feature selection, we concatenated all the features of the three modalities, then we performed feature selection and trained the classifier in the same way as for the single modalities.
		
	\subsection{Classification of Personality, PANAS and Social Context from Short and Long Videos}\label{PersonalityRecognition}
	
		\subsubsection{Single Modality Classification}\label{Class_long}
			\changemarker{For personality traits, PANAS and social context prediction, 7 scenarios have been tested. 
			The different scenarios have been selected to show how the different stimuli as well as their combination perform in the recognition tasks.
			The first 4 scenarios (Video-N1, Video-P1, Video-B1 and Video-U1 scenarios) consider only the samples of each of the 4 long videos for prediction. 
			The fifth (Short-videos scenario) considers only the samples of the 16 short videos together. 
			The sixth (Long-videos scenario) considers all the samples of the 4 long videos together.
			And the seventh (All-videos scenario) considers the samples of all the 20 videos (short and long).
			The concatenation of the features of all the samples of each scenario and each of the modalities (EEG, ECG and GSR), were associated to the labels of personality traits, PA, NA and social context dimensions. 
			The dimensionality of the feature vector of each scenario is different, for instance the Video N1 scenario with the EEG modality has a feature vector with dimensionality of 7560 features (72 samples $\times$ 105 features) for each participant. } 
			
			\changemarker{For each scenario and participant, 8 support vector machine (SVM) classifiers with linear kernel \cite{Cristianini1999} were trained, one for each of the 5 personality traits, 2 for mood dimensions of PA and NA and 1 for social context prediction. The labels for personality and mood dimensions are divided into high and low classes using the median value of each personality and mood dimensions as threshold. In the case of social context, if the participant was in a group during the long videos experiment it was considered as positive class and negative if it was in individual configuration. Note that social context prediction was not implemented for the Short-videos scenario simply because it is not applicable.}
			
			\changemarker{To test the method we use leave-one-participant-out cross-validation, in which, during training, principal components analysis (PCA) \cite{Jolliffe1986} is performed over the features of all the participants resulting in a reduction to 36 PCA channels. Next, inspired by \cite{Bins2001}, channels were selected by clustering them using Pearson correlation coefficient ($\rho$) as distance measure. This is done by ranking the PCA channels according to their Fisher's linear discriminant $J$ calculated for the training set over each channel with respect to the labels. Channels with $J<0.1$ are discarded. Next, the channel with the highest $J$ is selected. By calculating the $\rho$ coefficient between the selected channel and the remaining channels, redundant channels are removed by discarding channels with $\rho>0.5$. From the remaining channels the one with the highest $J$ is then selected and the process is continued until all the channels are either selected or discarded. With the selected PCA channels, an SVM with linear kernel is trained over the training set and tested over the test set. The regularization parameter $C$ of the linear SVM was empirically set to $0.25$.} 
		
		\subsubsection{Fusion of Modalities}\label{Fusion_long}
			\changemarker{In order to use complementary information from different modalities, decision level fusion of the three modalities (EEG, ECG and GSR) was implemented for each scenario.
			Following \cite{koelstra2013fusion}, a meta-classification of class labels (M-CLASS) was implemented in which a linear SVM classifier is trained over the probabilistic outputs of the training samples and the training labels. The trained classifier is then used to predict the label of the test sample.} 

	\subsection{Results and Discussion}\label{Results_long}
	
		In Table~\ref{Affect_Exp2}, the mean F1-scores (mean F1-score for both classes) over all participants, for classification of valence and arousal, using the Gaussian Na\"{i}ve Bayes classifier, are presented for the different modalities. Three scenarios are included, the first considers only the short videos experiment samples, the second the long videos experiment samples and the third all the samples of both experiments. Results for feature level fusion of the three modalities are also included. Random baseline results (analytically determined) obtained by assigning labels randomly are also included.

		\begin{table}[t]
			\centering
			\fontsize{7}{8}\selectfont
			\renewcommand{\arraystretch}{1.2}
			\caption{\label{Affect_Exp2} Mean F1-scores (mean F1-score for negative and positive class) over participants for recognition of Valence and Arousal. Bold values indicate whether the F1-score distribution over subjects is significantly higher than $0.5$ according to an independent one-sample t-test ($p<.01$). Analytical results for voting at random are shown.}
			\scalebox{0.9}[0.9]{
				\setlength{\tabcolsep}{2pt}
				\begin{tabular}{|l|l|l|l|l|l|l|}
					\hline
					\hline
					\multirow{2}{0.1\textwidth}{\centering Modality} &    \multicolumn{2}{c|}{Short} &  \multicolumn{2}{c|}{Long} &      \multicolumn{2}{c|}{All} \\
					\cline{2-7}
					&    \multicolumn{1}{c|}{Valence} & \multicolumn{1}{c|}{Arousal}  &    \multicolumn{1}{c|}{Valence} & \multicolumn{1}{c|}{Arousal} &    \multicolumn{1}{c|}{Valence} & \multicolumn{1}{c|}{Arousal} \\
					\hline\hline
					EEG		  & \textbf{0.576*}   & \textbf{0.592**}  &  \textbf{0.557**}  & \textbf{0.571**} & \textbf{0.564**} & \textbf{0.577**} \\
					\hline
					GSR 	  & 0.531   & 0.548     &  0.528    & \textbf{0.536*} & 0.528 & \textbf{0.541**} \\
					\hline
					ECG 	  & 0.535   & 0.550     &  \textbf{0.550**}  & \textbf{0.543*} & \textbf{0.545**} & \textbf{0.551**} \\
					\hline
					Fusion	  & \textbf{0.570*}   & \textbf{0.585**}  &  \textbf{0.551**}  & \textbf{0.569**} & \textbf{0.560**} & \textbf{0.564**} \\
					\hline\hline
					Random    & 0.500   & 0.500  & 0.500 & 0.500   & 0.500  & 0.500 \\
					\hline					
				\end{tabular}
			}
			\vspace{-.3cm}
		\end{table}
	
		Random levels for all the scenarios for valence and arousal had $0.5$ mean F1-score each. Significant higher than chance ($p<.01$ according to an independent one-sample t-test) F1-scores were obtained for all the scenarios using the EEG modality, for the long videos and all videos scenarios using ECG, and only for arousal recognition in the long videos and all videos scenarios using GSR. In general, arousal recognition got higher performance than valence, except for ECG modality in the long videos experiment. For all scenarios of valence and arousal recognition, EEG got significantly higher performance than ECG and GSR ($p<0.0001$ for both), resulting in a mean improvement, over the three scenarios, of $2.2\%$ and $3.2\%$ for recognition of valence and arousal over the ECG. ECG is still significantly better ($p<0.05$) than the GSR modality. Feature level fusion does not improve the results but they are still significantly higher than chance ($p<0.01$). Prediction of valence and arousal in short videos was better than in the long videos but the differences are not significant ($p=0.32$ for valence and $p=0.19$ for arousal). Results for recognition of valence and arousal using the videos of both experiments are not better than for each experiment alone. Our baseline results show average performance compared with the literature for recognition of valence and arousal \cite{koelstra2012deap,DECAF2015,SoleymaniMAHNOB2012}.

		\changemarker{In Table~\ref{Personality_Exp2}, the mean F1-score of the positive and negative classes over all participants for binary classification of personality traits, PANAS and social context is presented. In the table, the seven scenarios described in Sec.\ref{Class_long} are included. We have also implemented the baseline method proposed by Abadi et al \cite{FG_Paper}, based on a linear regression model for predictions using two physiological modalities, namely EEG and physiological signals (ECG+GSR). In \cite{FG_Paper}, they use only short videos and 35 participants. For the sake of comparison, we applied their method over the same 37 participants used in this study in the short videos experiment. Empirically estimated baseline results obtained by randomly assigning the labels according to the class ratio of the population are also reported.}
	
		\begin{table}[t]
			\centering
			\fontsize{7}{8}\selectfont
			\renewcommand{\arraystretch}{1.2}
			\caption{\label{Personality_Exp2} Mean F1-score (mean F1-score for negative and positive class) over participants, for personality traits (Extraversion, Agreeableness, Conscientiousness, Emotional Stability and Openness), PANAS (PA and NA) and social context recognition (number of 20-s segments stated in parenthesis). Bold values indicate whether the F1-score distribution over subjects is significantly higher than $0.5$ according to an independent one-sample t-test ($p<.001$). Results obtained with a baseline method \cite{FG_Paper}, for prediction of personality and PANAS using the short videos experiment are included for comparison. Empirical results for voting at random are also shown.}
			\scalebox{0.9}[0.9]{
				\setlength{\tabcolsep}{2pt}
				\begin{tabular}{|l|c|c|c|c|c|c|c|c|c|}
					\hline
					\hline
					\centering Scenario & Modality & \multicolumn{1}{c|}{Extr.}& \multicolumn{1}{c|}{Agre.}& \multicolumn{1}{c|}{Cons.}& \multicolumn{1}{c|}{Emot.}& \multicolumn{1}{c|}{Open.}& \multicolumn{1}{c|}{PA.}& \multicolumn{1}{c|}{NA.} & \multicolumn{1}{c|}{S. C.}\\
					\hline\hline
					\multirow{3}{*}{Video N1 (72)}   & EEG & \textbf{0.535} & 0.459 & \textbf{0.728} & \textbf{0.595} & 0.426 & \textbf{0.567} & 0.234 & 0.401 \\
					\cline{2-10}  & GSR & \textbf{0.675} & \textbf{0.699} & 0.284 & 0.405 & 0.459 & 0.431 & 0.327 & \textbf{0.644} \\
					\cline{2-10}  & ECG & 0.401 & 0.351 & \textbf{0.702} & \textbf{0.593} & \textbf{0.621} & 0.322 & 0.316 & 0.383 \\
					\hline\hline 
					\multirow{3}{*}{Video P1 (58)}  & EEG & \textbf{0.590} & 0.262 & 0.271 & 0.378 & \textbf{0.621} & \textbf{0.648} & \textbf{0.584} & \textbf{0.648} \\
					\cline{2-10}  & GSR & 0.485 & 0.162 & \textbf{0.649} & 0.405 & \textbf{0.756} & 0.401 & \textbf{0.648} & 0.405 \\
					\cline{2-10}  & ECG & 0.431 & 0.405 & \textbf{0.619} & \textbf{0.619} & 0.431 & \textbf{0.648} & \textbf{0.584} & 0.405 \\
					\hline\hline 
					\multirow{3}{*}{Video B1 (72)}   & EEG & \textbf{0.675} & \textbf{0.619} & \textbf{0.644} & 0.324 & 0.135 & 0.401 & \textbf{0.745} & 0.449 \\
					\cline{2-10}  & GSR & 0.316 & \textbf{0.730} & \textbf{0.728} & 0.473 & \textbf{0.648} & 0.322 & 0.251 & \textbf{0.539} \\
					\cline{2-10}  & ECG & \textbf{0.552} & \textbf{0.595} & \textbf{0.584} & \textbf{0.837} & 0.480 & \textbf{0.593} & \textbf{0.670} & 0.439 \\
					\hline\hline 
					\multirow{3}{*}{Video U1 (44)}  & EEG & 0.080 & 0.432 & 0.495 & \textbf{0.619} & 0.105 & \textbf{0.565} & \textbf{0.750} & 0.348 \\
					\cline{2-10}  & GSR & 0.431 & \textbf{0.675} & 0.348 & \textbf{0.730} & \textbf{0.560} & 0.485 & \textbf{0.598} & 0.401 \\
					\cline{2-10}  & ECG & 0.189 & 0.378 & \textbf{0.750} & \textbf{0.504} & 0.316 & \textbf{0.560} & \textbf{0.644} & \textbf{0.560} \\
					\hline\hline 
					\multirow{3}{*}{Short (94)}   & EEG & \textbf{0.730} & 0.351 & 0.347 & \textbf{0.567} & 0.486 & \textbf{0.565} & \textbf{0.598} & - \\
					\cline{2-10}  & GSR & 0.268 & \textbf{0.510} & \textbf{0.655} & 0.362 & \textbf{0.699} & 0.238 & 0.461 & - \\
					\cline{2-10}  & ECG & \textbf{0.621} & \textbf{0.513} & \textbf{0.590} & 0.140 & 0.483 & 0.426 & 0.362 & - \\
					\hline\hline 
					\multirow{3}{*}{Long (246)}  & EEG & \textbf{0.756} & 0.405 & 0.271 & \textbf{0.539} & 0.378 & 0.485 & \textbf{0.619} & \textbf{0.528} \\
					\cline{2-10}  & GSR & \textbf{0.567} & \textbf{0.674} & \textbf{0.539} & \textbf{0.565} & \textbf{0.782} & 0.485 & \textbf{0.584} & \textbf{0.835} \\
					\cline{2-10}  & ECG & \textbf{0.619} & 0.486 & 0.339 & \textbf{0.567} & 0.306 & 0.405 & 0.288 & \textbf{0.510} \\
					\hline\hline 
					\multirow{3}{*}{All (340)}  & EEG & 0.135 & \textbf{0.648} & 0.485 & 0.270 & 0.401 & \textbf{0.674} & 0.405 & 0.456 \\
					\cline{2-10}  & GSR & 0.371 & \textbf{0.837} & \textbf{0.535} & \textbf{0.621} & 0.371 & \textbf{0.649} & \textbf{0.547} & \textbf{0.702} \\
					\cline{2-10}  & ECG & 0.485 & \textbf{0.567} & 0.449 & 0.189 & \textbf{0.648} & 0.459 & \textbf{0.590} & \textbf{0.728} \\
					\hline\hline 
					\multicolumn{1}{|l|}{\multirow{1}{0.10\textwidth}{\cite{FG_Paper} Abadi et al}}  & EEG & 0.410 & 0.480 & 0.500 & 0.510 & \textbf{0.600} & 0.460 & 0.360 & - \\
					\cline{2-10}\multicolumn{1}{|l|}{\multirow{1}{0.11\textwidth}{\cite{FG_Paper} Abadi et al}}  & ECG+GSR & \textbf{0.670} & \textbf{0.570} & 0.530 & \textbf{0.640} & 0.500 & 0.500 & \textbf{0.560} & -  \\
					\hline\hline
					Random  & - &0.500  & 0.500   & 0.500  & 0.500 & 0.500   & 0.500  & 0.500  & 0.500  \\
					\hline
				\end{tabular}
			}
			\vspace{-.3cm}
		\end{table}
		
		\changemarker{Random mean F1-score is $0.5$ for all the scenarios and dimensions (personality traits, PANAS and social context). Different significant ($p<0.001$) F1-scores are observed for all the scenarios. Single long videos (Video-N1, Video-P1, Video-B1 and Video-U1 scenarios) show to be relevant for the prediction of different personality traits. Consistent significant results over the three modalities are observed for NA prediction in the Video-P1 and Video-U1 scenarios; for agreeableness and consciousness in the Video-B1 scenario; and emotional stability in the Video-U1 scenario. When considering the Short-videos scenario various modalities show contrasting performance. In the Long-videos scenario, consistent significant results are obtained for extroversion, emotional stability and social context. In this scenario, the GSR modality shows the best performance on average for the different dimensions than all other modalities and scenarios with a mean F1-score of $0.623$. In the All-videos scenario, only agreeableness gets consistent performance over each of the modalities.}  

		\changemarker{In comparison with the baseline method \cite{FG_Paper}, using only the short videos with the EEG modality, our method outperforms \cite{FG_Paper} in prediction of extroversion, emotional stability, PA and NA. It is interesting to note that both methods seem to work complementary to each other. Both methods fail to predict agreeableness and conscientiousness from EEG. Using physiological signals (ECG and GSR), our method outperforms \cite{FG_Paper} in prediction of conscientiousness and openness using the GSR and in prediction of conscientiousness using ECG. Considering the GSR modality of the Long-videos scenarios, our method outperforms \cite{FG_Paper} in prediction of agreeableness, conscientiousness, openness and NA.}
		
		\changemarker{Table~\ref{Pers_Exp1} presents the mean F1-score over all participants for binary classification of personality traits, PANAS and social context, for the decision level fusion scheme described in \ref{Fusion_long}. The same scenarios as for the single modality experiments are included. The results of the best performing single modalities for each scenario are also included.}
		
		\begin{table}[t]
			\centering
			\fontsize{7}{8}\selectfont
			\renewcommand{\arraystretch}{1.2}
			\caption{\label{Pers_Exp1} Mean F1-score (mean F1-score for negative and positive class) over participants, for recognition of personality traits, PANAS and social context, for fusion of modalities (See~\ref{Fusion_long}). Bold values indicate whether the F1-score distribution over subjects is significantly higher than $0.5$ according to an independent one-sample t-test ($p<.001$). The best performing single modality is also included.}
			\scalebox{0.9}[0.9]{
				\setlength{\tabcolsep}{2pt}
				\begin{tabular}{|l|c|c|c|c|c|c|c|c|c|}
					\hline
					\hline
					\centering Scenario & Fusion & \multicolumn{1}{c|}{Extr.}& \multicolumn{1}{c|}{Agre.}& \multicolumn{1}{c|}{Cons.}& \multicolumn{1}{c|}{Emot.}& \multicolumn{1}{c|}{Open.}& \multicolumn{1}{c|}{PA.}& \multicolumn{1}{c|}{NA.} & \multicolumn{1}{c|}{S. C.}\\
					\hline\hline
					\multirow{2}{*}{Video N1}  & M-CLASS  & 0.431 & 0.485 & \textbf{0.513} & \textbf{0.539} & 0.377 & 0.431 & 0.178 & \textbf{0.510} \\
					\cline{2-10}  & Best single modality & \textbf{0.675} & \textbf{0.699} & \textbf{0.728} & \textbf{0.595} & \textbf{0.621} & \textbf{0.567} & 0.327 & \textbf{0.644} \\
					\hline\hline 
					\multirow{2}{*}{Video P1}  & M-CLASS  & 0.431 & 0.135 & \textbf{0.510} & 0.432 & \textbf{0.675} & \textbf{0.621} & \textbf{0.699} & 0.431 \\
					\cline{2-10}  & Best single modality & \textbf{0.590} & 0.405 & \textbf{0.649} & \textbf{0.619} & \textbf{0.756} & \textbf{0.648} & \textbf{0.648} & \textbf{0.648} \\
					\hline\hline 
					\multirow{2}{*}{Video B1}  & M-CLASS  & \textbf{0.535} & \textbf{0.728} & \textbf{0.674} & \textbf{0.695} & 0.405 & 0.324 & \textbf{0.552} & 0.426 \\
					\cline{2-10}  & Best single modality & \textbf{0.675} & \textbf{0.730} & \textbf{0.728} & \textbf{0.837} & \textbf{0.648} & \textbf{0.593} & \textbf{0.745} & \textbf{0.539} \\
					\hline\hline 
					\multirow{2}{*}{Video U1}  & M-CLASS  & 0.162 & 0.459 & \textbf{0.584} & \textbf{0.730} & 0.322 & \textbf{0.615} & \textbf{0.770} & 0.348 \\
					\cline{2-10}  & Best single modality & 0.431 & \textbf{0.675} & \textbf{0.750} & \textbf{0.730} & \textbf{0.560} & \textbf{0.565} & \textbf{0.750} & \textbf{0.560} \\
					\hline\hline 
					\multirow{2}{*}{Short}  & M-CLASS  & \textbf{0.649} & 0.459 & \textbf{0.560} & 0.405 & \textbf{0.567} & 0.362 & \textbf{0.540} & - \\
					\cline{2-10}  & Best single modality & \textbf{0.730} & \textbf{0.513} & \textbf{0.655} & \textbf{0.567} & \textbf{0.699} & \textbf{0.565} & \textbf{0.598} & - \\
					\hline\hline 
					\multirow{2}{*}{Long}  & M-CLASS  & \textbf{0.648} & \textbf{0.510} & 0.268 & \textbf{0.513} & \textbf{0.535} & 0.449 & \textbf{0.699} & \textbf{0.725} \\
					\cline{2-10}  & Best single modality & \textbf{0.756} & \textbf{0.674} & \textbf{0.539} & \textbf{0.567} & \textbf{0.782} & 0.485 & \textbf{0.619} & \textbf{0.835} \\
					\hline\hline 
					\multirow{2}{*}{All} & M-CLASS  & 0.297 & \textbf{0.703} & 0.401 & 0.459 & 0.417 & \textbf{0.644} & 0.446 & \textbf{0.648} \\
					\cline{2-10}  & Best single modality & 0.485 & \textbf{0.837} & \textbf{0.535} & \textbf{0.621} & \textbf{0.648} & \textbf{0.674} & \textbf{0.590} & \textbf{0.728} \\
					\hline
				\end{tabular}
			}
		\vspace{-.3cm}
		\end{table}
		
		\changemarker{We can see from Table~\ref{Pers_Exp1} that feature level fusion only outperformed the best single modality in a few cases. The difference is only significant for prediction of NA in the Video-P1 and Long-videos scenarios and for prediction of PA in the Video-U1 scenario. In the remaining cases, the weakest modalities seem to undermine the performance of the best modality, but still it is possible to predict conscientiousness and NA in 5 scenarios. It is interesting to note that, though individual long videos do not perform well for social context prediction, using the samples of the 4 long videos experiment together (Long-videos scenario) performs relatively well with mean F1-score of $0.725$. The All-videos scenario which includes samples of both short and long videos does not lead to better performance.}
			
		\changemarker{We believe that these results can be improved by the use of different feature extraction and selection methods, such as deep belief networks. We encourage researchers to try and use this challenging dataset.} 
		

\section{Conclusions}\label{conclusions}

	In this work, we presented a dataset for multimodal research of affect, personality traits and mood on individuals and groups by means of neuro-physiological signals. 
	We found significant correlations between internal and external affect annotations of valence and arousal, indicating that external annotation is a good predictor of the affective state of participants. We showed that social context has an important effect on the valence and arousal expressed by the participants, given that group participants showed lower levels of arousal for low arousal clips, and higher levels of arousal for high arousal clips and in general higher valence than when they are alone. PA showed to be significantly correlated with arousal expressed during long videos. 
	EEG was the best modality for prediction of valence and arousal, while feature level fusion did not improve the results. 
	\changemarker{For prediction of personality traits, PANAS and social context, GSR of long videos is the best modality over all dimensions with a mean F1-score of $0.623$. Finally, feature level fusion improved the results for NA and PA prediction. The database is publicly available.}

\section{Acknowledgments}
The first author acknowledges support from CONACyT, Mexico, through a scholarship to pursue graduate studies at Queen Mary University of London.

%


%





\ifCLASSOPTIONcaptionsoff
  \newpage
\fi



%

\bibliographystyle{ieeetran}
\bibliography{Paper_TAC}  

%

\begin{IEEEbiography}[{\includegraphics[width=1in,height=1.25in,clip,keepaspectratio]{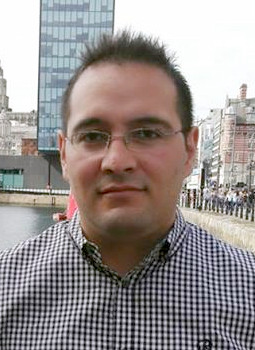}}]{Juan Abdon Miranda Correa}
	received the MSc degree in electronics systems from Tecnol\'{o}gico de Monterrey, Campus Toluca, Mexico, in 2012. He is now working towards the PhD degree at the School of Electronic Engineering and Computer Science, Queen Mary University of London, UK. His research interests include: multimodal affect recognition in human computer interaction, analysis of social interaction in affective multimedia and deep learning.

\end{IEEEbiography}

\begin{IEEEbiography}[{\includegraphics[width=1in,height=1.25in,clip,keepaspectratio]{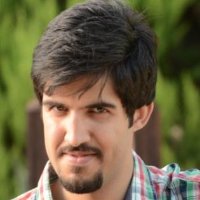}}]{Mojtaba Khomami Abadi}
	is a PhD candidate at the Department of Information Engineering and Computer Science, University of Trento, Italy. Mojtaba is also the CTO of Sensaura Inc., a Canadian startup on real-time and multimodal emotion recognition technologies. His research interests include: user centric affective computing in human computer interaction and affective multimedia analysis.
\end{IEEEbiography}


\begin{IEEEbiography}[{\includegraphics[width=1in,height=1.25in,clip,keepaspectratio]{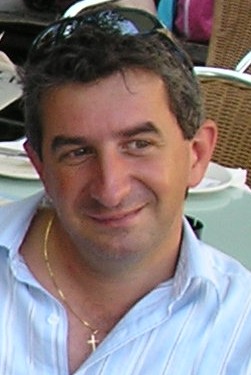}}]{Nicu Sebe}
	received the PhD degree from Leiden University, The Netherlands, in 2001. Currently, he is with the Department of Information Engineering and Computer Science, University of Trento, Italy, where he is leading the research in the areas of multimedia information retrieval and human behavior understanding. He was a general co-chair of FG 2008 and ACM Multimedia 2013, and a program chair of CIVR 2007 and 2010, and ACM Multimedia 2007 and 2011. He is a program chair of ECCV 2016 and ICCV 2017. He is a senior member of the IEEE and ACM and a fellow of IAPR.
\end{IEEEbiography}


\begin{IEEEbiography}[{\includegraphics[width=1in,height=1.25in,clip,keepaspectratio]{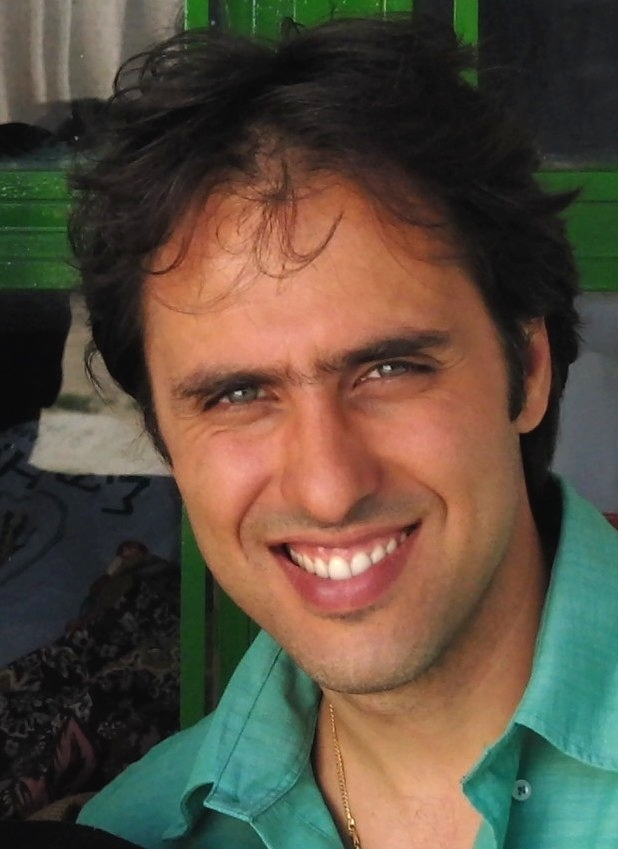}}]{Ioannis Patras}
	received the PhD degree from the Delft University of Technology, The Netherlands, in 2001. He is a senior lecturer in computer vision in Queen Mary, University of London. He was in the organizing committee of IEEE SMC2004, FGR2008, ICMR2011, ACMMM2013 and was the general chair of WIAMIS2009. His research interests include computer vision, pattern recognition and multimodal HCI. He is a senior member of the IEEE.
\end{IEEEbiography}




\end{document}